\documentclass[%
 reprint,
superscriptaddress,
 amsmath,amssymb,
 aps,
]{revtex4-2}

\usepackage{physics}
\usepackage{graphicx}
\usepackage{dcolumn}
\usepackage{bm}
\usepackage{xcolor}
\usepackage{setspace}
\usepackage[utf8]{inputenc} 
\DeclareUnicodeCharacter{03B2}{$\beta$}

\usepackage{hyperref}
\hypersetup{
    colorlinks=true,   
    citecolor=blue,    
    linkcolor=black,    
    urlcolor=black,    
    anchorcolor=black   
}

\usepackage{titlesec}
\titleformat{\section}
  {\normalfont\bfseries\centering}
  {\thesection}{1em}{}

\begin{document}

\newcommand{\jagang}[1]{{\color{purple} #1}}
\newcommand{\blue}[1]{{\color{blue} #1}}
\preprint{APS/123-QED}

\title{Spontaneous emission decay and excitation in photonic time crystals}

\author{Jagang Park}
\thanks{These authors contributed equally to this work.}
\affiliation{%
 Department of Electrical Engineering and Computer Sciences, University of California, Berkeley, California 94720, USA
}%

\author{Kyungmin Lee}
\thanks{These authors contributed equally to this work.}
\affiliation{%
 Department of Physics, Korea Advanced Institute of Science and Technology, Daejeon 34141, Republic of Korea
}%

\author{Ruo-Yang Zhang}
\affiliation{%
 Department of Physics, the Hong Kong University of Science and Technology, Clear Water Bay, Kowloon, Hong Kong 999077, China
}%

\author{Hee-Chul Park}
\affiliation{%
Department of Physics, Pukyong National University, Busan 48513, Republic of Korea 
}%

\author{Jung-Wan Ryu}
\affiliation{%
 Center for Theoretical Physics of Complex Systems, Institute for Basic Science, Daejeon 34126, Republic of Korea
}%

\author{Gil Young Cho}
\affiliation{%
 Department of Physics, Pohang University of Science and Technology, Pohang 37673, Republic of Korea
}%

\author{Min Yeul Lee}
\affiliation{%
 Department of Material Sciences and Engineering, Korea Advanced Institute of Science and Technology, Daejeon 34141, Republic of Korea
}%

\author{Zhaoqing Zhang}
\affiliation{%
 Department of Physics, the Hong Kong University of Science and Technology, Clear Water Bay, Kowloon, Hong Kong 999077, China
}%

\author{Namkyoo Park}
\affiliation{%
 Department of Electrical and Computer Engineering, Seoul National University, Seoul 08826, Republic of Korea
}%

\author{Wonju Jeon}
\affiliation{%
 Department of Mechanical Engineering, Korea Advanced Institute of Science and Technology, Daejeon 34141, Republic of Korea
}%

\author{Jonghwa Shin}
\affiliation{%
 Department of Material Sciences and Engineering, Korea Advanced Institute of Science and Technology, Daejeon 34141, Republic of Korea
}%

\author{C. T. Chan}
\affiliation{%
 Department of Physics, the Hong Kong University of Science and Technology, Clear Water Bay, Kowloon, Hong Kong 999077, China
}%

\author{Bumki Min}
  \email{bmin@kaist.ac.kr}
\affiliation{%
 Department of Physics, Korea Advanced Institute of Science and Technology, Daejeon 34141, Republic of Korea
}%







\begin{abstract} 
Over the last few decades, the predominant strategies for controlling spontaneous emission have involved tailoring the spatial surroundings of quantum emitters or atoms to create resonant or spatially periodic photonic structures. However, the rise of time-varying photonics has prompted a reevaluation of spontaneous emission in dynamically changing environments, especially within photonic time crystals, where optical properties undergo time-periodic modulation. Here, we apply classical light–matter interaction theory together with Floquet analysis to reveal a substantial enhancement of the spontaneous emission decay rate at the momentum gap frequency in photonic time crystals. Moreover, our findings suggest that photonic time crystals enable a non-equilibrium light–matter interaction process: the spontaneous excitation of an atom from its ground state to an excited state, accompanied by the concurrent emission of a photon, referred to as spontaneous emission excitation.


\end{abstract}

\maketitle
Investigation of electromagnetic wave dynamics in space-time periodic media began in the 1950s \cite{morgenthaler1958velocity,1138637,Cassedy1967,fante1971transmission,felsen1970wave}, initially focusing on temporally growing instabilities in distributed parametric media. However, time-varying photonics gained widespread attention much later. A pioneering experiment with a time-periodic transmission line revealed a shallow yet definitive momentum gap, marking a key milestone despite limited initial recognition \cite{doi:10.1063/1.4928659}. This finding spurred the extension of photonic crystals and metamaterials into the space-time domain, leveraging the additional temporal degree of freedom for enhanced dispersion and band structure engineering \cite{Chamanara2018,Shcherbakov2019,PhysRevLett.123.206101,pacheco2020effective,Park2021,Lee2021}. Advances in time-varying photonics have since enabled exploration of broadband nonreciprocity \cite{Sounas2017}, one-way amplification \cite{PhysRevLett.123.206101}, parametric oscillation \cite{doi:10.1126/sciadv.adg7541,doi:10.1126/sciadv.abo6220}, pulse compression \cite{8889149}, harmonic generation \cite{Shcherbakov2019}, and even Hawking radiation mimicry \cite{doi:10.1073/pnas.2302652120}.

Only recently has the Floquet‑system framework been refined, deepening our understanding of photonic time crystals (PTCs), which are physically distinct from Wilczek’s many‑body \textit{time crystal} phases \cite{wilczek2012quantum,shapere2012classical}. By long‑standing convention, however, the same class of time‑periodic photonic media has been referred to as a \textit{temporal photonic crystal} \cite{zurita2009reflection,doi:10.1063/1.4928659,salem2015temporal,PhysRevA.93.063813,martinez2018parametric}. The effective Hamiltonian matrix for PTCs, derived from Maxwell's equations, is identified as pseudo-Hermitian, which leads to the photonic Floquet eigenmodes becoming non-orthogonal. Furthermore, the momentum gap has been proven to be the phase in which pseudo-Hermiticity is broken along the wavenumber axis, with its edges characterized as exceptional points (EPs) \cite{PhysRevB.98.085142,doi:10.1126/sciadv.abo6220,wang2019non}. These findings highlight the need for a non-Hermitian theoretical framework to analyze classical PTCs, whose non-Hermitian dynamics strongly shape their light–matter interactions. Considering these non-Hermitian dynamics, we show that the spontaneous emission decay rate at the momentum gap frequency in PTCs is significantly enhanced, contradicting a recent study \cite{doi:10.1126/science.abo3324} that predicted its complete vanishing.

To accurately quantify light-matter interactions in PTCs, we must account for time-periodicity-induced loss and gain regions in wavenumber-frequency space, reflected in positive or negative momentum-resolved photonic density of states ($\mathrm{kDOS}$). The negative $\mathrm{kDOS}$ observed in the gain region necessitates a gain-induced correction, prompting a re-evaluation of the spontaneous emission decay rate \cite{PhysRevLett.127.013602, PhysRevA.109.013513, ren2021quasinormal, vyshnevyy2022gain, scarlatella2019spectral}. Intriguingly, gain in PTCs can trigger spontaneous excitation of an atom, accompanied by photon emission, highlighting the rich spectrum of light-matter interactions available in nonequilibrium photonics. Additionally, we show that the non-orthogonality between photonic Floquet eigenmodes, quantified by the Petermann factor ($\mathrm{PF}$)  \cite{1070064,Berry2003,Zhang2018,wang2020petermann}, increases both the spontaneous emission decay and excitation rates.

\begin{figure}[htb!]
  \centering
    \includegraphics[width=0.45\textwidth]{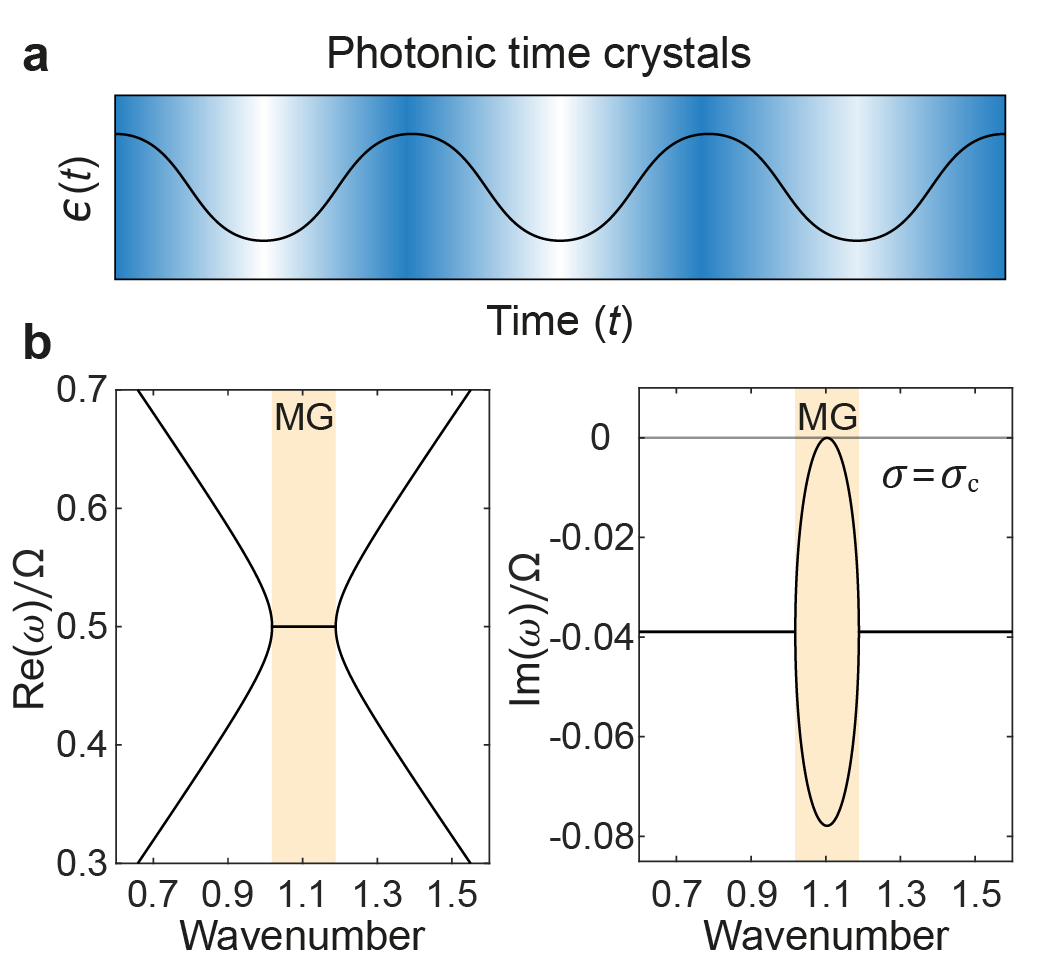}
  \caption{\label{fig:Fig1} (a) Time periodic permittivity modulation. (b) Floquet band structure of a photonic time crystal with $\epsilon(t)=\epsilon_0+\epsilon_m \sin{\Omega t}$. The plot uses $\epsilon_0=5$, $\epsilon_m=1.5$, $\mu=1$, and $\Omega=1$, with conductivity set to the critical value $\sigma=\sigma_c = 0.3715$. The momentum gap (MG) is highlighted in yellow.}
\end{figure}

In this Letter, we employ classical light-matter interaction theory, treating a quantum emitter as an oscillating dipole while the electromagnetic field obeys Maxwell’s equations.  By modeling the emitter as an oscillating point dipole, $\mathbf{p}(\mathbf{r},t)=\mathrm{Re}[\mathbf{p}\delta(\mathbf{r}-\mathbf{r}_0)\exp(-i\omega t)]$, we recast the inherently quantum interaction in a tractable classical framework \cite{Milonni1975,777216}. Notably, the spontaneous emission decay rate in the weak coupling regime can be estimated from classical power flow or radiation reaction arguments, even in linearly amplifying media \cite{PhysRevLett.127.013602,PhysRevA.109.013513,ren2021quasinormal}. Hence, the problem can be analyzed by solving Maxwell’s equations for a medium with time-periodic permittivity, $ \epsilon(t) = \epsilon(t + 2\pi/\Omega) $, where $ \Omega $ is the modulation frequency (Fig. \ref{fig:Fig1}a):
\begin{align}
    \nabla\times\mathbf{E}(\mathbf{r},t) &= -\mu\frac{\partial}{\partial t}\mathbf{H}(\mathbf{r},t), \label{eqn:Maxwell1} \\
    \nabla\times\mathbf{H}(\mathbf{r},t) &= \frac{\partial}{\partial t}[\epsilon(t)\mathbf{E}(\mathbf{r},t)]+\mathbf{J}(\mathbf{r},t) +\sigma\mathbf{E}(\mathbf{r},t).\label{eqn:Maxwell2} 
\end{align}
Here, vacuum permittivity and permeability are set to unity. The conductivity $ \sigma $ accounts for intrinsic loss ($ \sigma > 0 $) in light-matter interactions, although other loss models are possible \cite{PhysRevB.98.085142} (see Supplemental Material D \cite{SM}). We include the dipole-induced current source $\mathbf J=\partial\mathbf p/\partial t$ in Maxwell’s equations for the PTC. Combining Eqs. \eqref{eqn:Maxwell1} and \eqref{eqn:Maxwell2} yields a single differential equation that can be cast as an eigenvalue problem with a Floquet Hamiltonian $\mathcal H_F$ (see Supplemental Material A \cite{SM}).

To obtain the photonic Floquet band structure, we solve the momentum-resolved eigenvalue problem. Fig.\ref{fig:Fig1}b shows the resulting bands and highlights the momentum gap, where the imaginary part of the eigenfrequency bifurcates. For a sinusoidally modulated permittivity,
\(\epsilon(t)=\epsilon_{0}+\epsilon_{m}\sin\Omega t\), the conductivity \(\sigma\) alone determines the maximum positive imaginary part of the eigenfrequency within the gap, denoted \(\gamma_{\mathrm{max}}\). We define the critical conductivity $\sigma_c$ as the value where $\gamma_{\mathrm{max}}=0$. When $\sigma<\sigma_c$ ($\gamma_{\mathrm{max}}>0$) the system lies in the small intrinsic-loss regime, whereas $\sigma>\sigma_c$ keeps all imaginary parts negative, corresponding to the large intrinsic-loss regime.


Quantifying the spontaneous emission decay rate in a PTC begins with the time-averaged power radiated by a point dipole. In spatial Fourier space, this power equals the momentum integral of the contributions from extended dipole components (see Supplemental Material B \cite{SM}):
\begin{equation}
    \begin{aligned}
    \bar{P}(\omega) &= \int_{\mathbb{K}}{\bar{P}(\mathbf{k},\omega)\ d^3\mathbf{k}}\\
    &= \frac{1}{(2\pi)^3}\frac{\pi\omega^2|\mathbf{p}|^2}{4\epsilon_0}\int_{\mathbb{K}}{\rho_\mathbf{p}(\mathbf{k},\omega)\ d^3\mathbf{k}}.
\end{aligned}
\end{equation}
Here, $\rho_\mathbf{p}(\mathbf{k}, \omega)$ is the momentum-resolved photonic density of states ($\mathrm{kDOS}$) projected onto the dipole orientation,  
\begin{equation}
\rho_\mathbf{p}(\mathbf{k},\omega)\equiv \frac{2\epsilon_0\mu\omega}{\pi }\mathrm{Im}[\mathbf{n}_\mathbf{p}\cdot\mathbb{G}_0(\mathbf{k},\omega)\cdot\mathbf{n}_\mathbf{p}],\label{eqn:LDOSww'}
\end{equation}
with $\mathbf{n}_\mathbf{p}= |\mathbf{p}|/\mathbf{p}$ and $\mathbb{G}_0$ the dyadic Green's function. In this formulation, $\mathbb{G}_0$ is a projection of the full Green's function of the Floquet Hamiltonian matrix, weighted by the Fourier components of $1/\epsilon(t)$,
\begin{equation}
    \begin{aligned}
    \mathcal{G}_F(\mathbf{k},\omega)
    &= [\omega\mathbf{I}_F-\mathcal{H}_F(\mathbf{k})]^{-1}\\
    &=\sum_{m}\frac{1}{\omega-\omega_{m}(\mathbf{k})}\frac{\ket{R_{m}(\mathbf{k})}\bra{L_{m}(\mathbf{k})}}{\braket{L_{m}(\mathbf{k})}{R_{m}(\mathbf{k})}}
\end{aligned}
\end{equation}
evaluated at the excitation frequency $\omega$. Here, $m$ indexes the modes, $\omega_m=\Omega_m-i\gamma_m$ are the complex quasi-eigenfrequencies, and $\bra{L_m}$ and $\ket{R_m}$ are the left and right eigenvectors of $\mathcal{H}_F$ (see Supplementary Material A \cite{SM}). 

\begin{figure*}[htb!]
  \centering
    \includegraphics[width=0.9\textwidth]{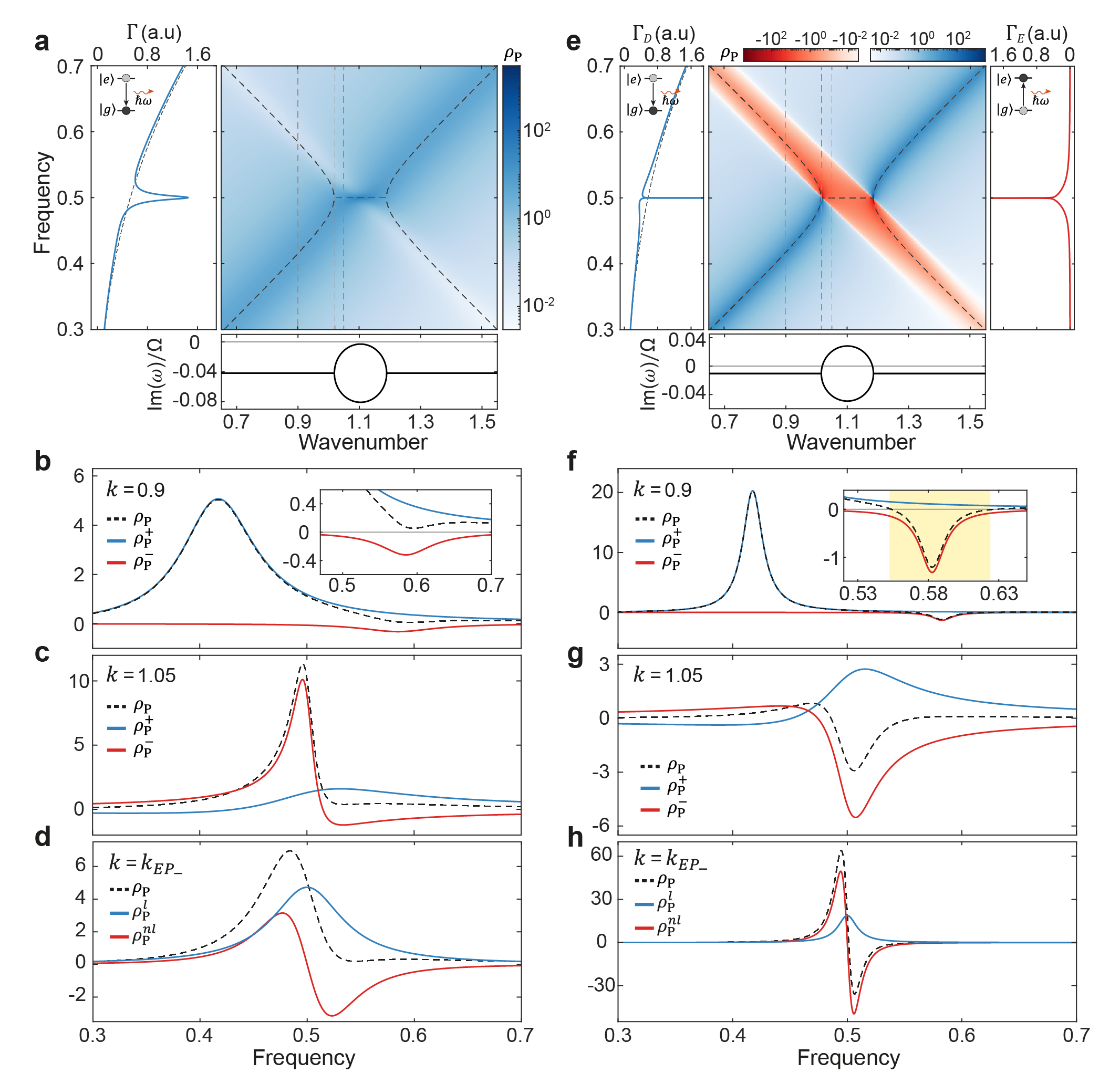}
  \caption{\label{fig:Fig2}(a) Map of the kDOS, the imaginary part of the quasi-eigenfrequency, and the spontaneous emission decay rate $\Gamma$ for $\sigma = 0.4 > \sigma_{c}$ (large-loss case). (b–d) Plots of the kDOS and its modal decompositions versus frequency for $k = 0.9$ (outside the momentum gap), $k = 1.05$ (within the gap), and $k = k_{EP_-}$. Outside the gap, the kDOS is well represented by the sum of two weighted, symmetric Lorentzian profiles of opposite sign, whereas inside the gap it becomes asymmetric. At $k = k_{EP_-}$, the kDOS decomposes into a symmetric Lorentzian and a term proportional to its frequency derivative. (e) Map of the kDOS, the imaginary part of the quasi-eigenfrequency, and the spontaneous emission decay and excitation rates for $\sigma = 0.1 < \sigma_{c}$ (small loss case). (f–h) Corresponding plots of the kDOS and its decompositions at $k = 0.9$, $1.05$, and $k_{EP_-}$. Outside the gap, the kDOS becomes negative near the negative-frequency Floquet sideband, and this negative region extends into the central gap. All calculations assume $\mathbf k \perp \mathbf n_{\mathbf p}$ except for the decay and excitation curves. The spontaneous emission decay rate in a homogeneous, time-invariant medium with permittivity $\epsilon_{0}$ is plotted as a black dashed line in (a) and (e).}
\end{figure*}


Focusing on the two modes $\omega_\pm=\Omega_\pm -i\gamma_\pm$, which form the momentum gap at $\omega=\Omega/2$, and assuming $\mathbf{k}\perp\mathbf{n}_\mathbf{p}$, the $\mathrm{kDOS}$ can be approximated as follows (for $\mathbf{k}\parallel\mathbf{n}_\mathbf{p}$, see Supplemental Material B \cite{SM}):  
\begin{equation}
    \begin{aligned}
    \rho_\mathbf{p}(\mathbf{k},\omega) &\approx \sum_{\alpha=\pm} \rho_\mathbf{p}^\alpha(\mathbf{k},\omega)\\
    &\approx\sum_{\alpha=\pm}\frac{1}{\pi}\frac{\gamma_\alpha}{(\omega-\Omega_\alpha )^2+\gamma_\alpha^2}\mathrm{Re}[I_\alpha (\mathbf{k})]\\
    &\quad+\frac{1}{\pi}\frac{\Omega_\alpha-\omega}{(\omega-\Omega_\alpha)^2+\gamma_\alpha^2}\mathrm{Im}[I_\alpha (\mathbf{k})].
\end{aligned}
\end{equation}
Here, $I_\alpha(\mathbf{k})$ is the complex, normalized field intensity projected onto the dipole; its real and imaginary parts weight the absorptive (symmetric) and dispersive (antisymmetric) Lorentzian terms, respectively. Far from the momentum gap, the imaginary part of $I_{\alpha}(\mathbf{k})$ is negligible compared with its real part. In addition, $\mathrm{Re}[I_+(\mathbf{k})]$ is positive, whereas $\mathrm{Re}[I_-(\mathbf{k})]$ is negative. Hence, at any fixed wavenumber far from the gap, the $\mathrm{kDOS}$ reduces to the sum of two weighted, symmetric Lorentzian profiles of opposite sign. Each symmetric Lorentzian is weighted by $\mathrm{Re}[I_{\alpha}(\mathbf{k})]$, which grows sharply as the wavenumber approaches the edge from within the band. It is also worth noting that, in the time-invariant limit ($\epsilon_m \rightarrow 0$), $\rho_\mathbf{p}^-(\mathbf{k},\omega)$ vanishes, and the $\mathrm{kDOS}$ reduces to that of the unmodulated medium.

We first consider the large intrinsic loss case, where $\sigma>\sigma_c$ (equivalently, $\gamma_{\mathrm{max}}<0$). In Fig. \ref{fig:Fig2}a, the $\mathrm{kDOS}$ map remains positive throughout the entire $(\mathbf{k}, \omega)$ plane. Figures \ref{fig:Fig2}b and \ref{fig:Fig2}c show the $\mathrm{kDOS}$ and its modal decompositions at $k=0.9$ (outside the gap) and $k=1.05$ (within the gap), respectively. Outside the gap, the $\mathrm{kDOS}$ attains a shallow minimum near the negative-frequency Floquet sideband owing to the negative contribution from $\rho_\mathbf{p}^-$, yet it stays positive overall. Within the gap, $\mathrm{Im}[I_{\alpha}(\mathbf{k})]$ becomes appreciable, so $\rho_\mathbf{p}^\alpha(\mathbf{k},\omega)$ departs from a purely symmetric Lorentzian (Fig. \ref{fig:Fig2}c). At $k=k_{EP_-}$ (the left edge of the gap), the $\mathrm{kDOS}$ is well approximated by the sum of a symmetric Lorentzian and a term proportional to its frequency derivative (Fig. \ref{fig:Fig2}d, see Supplemental Material G \cite{SM}). In the large intrinsic loss regime, the $\mathrm{kDOS}$ spontaneous emission model remains valid, predicting an enhanced yet finite spontaneous emission decay rate near the momentum gap frequency (left panel of Fig. \ref{fig:Fig2}a). For reference, the spontaneous emission decay rate in a homogeneous, time-invariant medium with permittivity $\epsilon_0$ is shown as a black dashed line.

We now turn to the low loss regime, defined by \(\sigma<\sigma_c\) (equivalently, \(\gamma_{\mathrm{max}}>0\)). In this case, the $\mathrm{kDOS}$ outside the gap becomes negative near the negative-frequency Floquet sideband (Figs.~\ref{fig:Fig2}e,f) and the negative region extends into the central gap (Figs.~\ref{fig:Fig2}f–h). Negative local densities of states have been considered in spontaneous emission from linear gain media \cite{ren2021quasinormal,vyshnevyy2022gain} and in the spectral functions of driven dissipative quantum systems \cite{scarlatella2019spectral}.  Here, the sign reversal arises from the net gain introduced by the time periodic permittivity modulation; Supplemental Material~D \cite{SM} shows the same effect in a driven Lorentz oscillator with a time periodic spring constant. To evaluate the spontaneous emission decay rate, we calculate the time-averaged power radiated into the momentum domain \(\mathbb{K}_D\) where the \(k\)DOS is positive,
\begin{equation}
\label{eqn:SED_rate}
\bar{P}_D(\omega)=\frac{1}{(2\pi)^3}\,
\frac{\pi\omega^{2}|\mathbf{p}|^{2}}{4\epsilon_{0}}
\int_{\mathbb{K}_D}\rho_{\mathbf{p}}(\mathbf{k},\omega)\,d^{3}\mathbf{k},
\end{equation}
and compare it with the vacuum reference \(P_0(\omega)\). The corrected decay rate then follows as \(\Gamma_D(\omega)=\Gamma_0(\omega)\,\bar{P}_D(\omega)/P_0(\omega)\) \cite{ren2021quasinormal,PhysRevA.109.013513}.

Within a quantum framework, negative $\mathrm{kDOS}$ implies a spontaneous transition of a two-level atom from its ground to excited state \cite{PhysRevLett.127.013602}, accompanied by photon emission, which we refer as \textit{spontaneous emission excitation}. In regions with a negative kDOS, the radiation reaction acting on an extended dipole source exhibits negative values, indicative of \textit{negative damping}. We interpret this as a spontaneous emission process accompanied by the excitation of an atom. In a time-invariant medium, spontaneous emission excitation is forbidden, because energy conservation prevents an atom from gaining internal energy while emitting a photon when no external modulation is present. A PTC, however, makes this process possible: the temporal modulation supply discrete quanta that exactly match the energy deficit, thereby enabling the atom to draw energy from the modulation, emit a photon, and end in an excited state. The corresponding spontaneous emission excitation rate is given by $\Gamma_E(\omega)=\Gamma_0(\omega)\bar{P}_E(\omega)/P_0(\omega)$, where the time-averaged power absorbed by the dipole is 
\begin{equation}\label{eqn:SEE_rate}
    \bar{P}_E(\omega) = \frac{1}{(2\pi)^3} \frac{\pi\omega^2|\mathbf{p}|^2}{4\epsilon_0} \int_{\mathbb{K}_E}{|\rho_\mathbf{p}(\mathbf{k},\omega_0)|\ d^3\mathbf{k}},
\end{equation}
with the integral restricted to the momentum domain $\mathbb{K}_E$ where the $\mathrm{kDOS}$ is negative. Figure \ref{fig:Fig2}e presents the $\mathrm{kDOS}$ map for the small intrinsic loss case, with the spontaneous emission decay rate shown in the left panel and the excitation rate in the right panel. The decomposition of the $\mathrm{kDOS}$ within the gap is provided in Supplemental Material E \cite{SM}.

\begin{figure}[h!]
  \centering
    \includegraphics[width=0.45\textwidth]{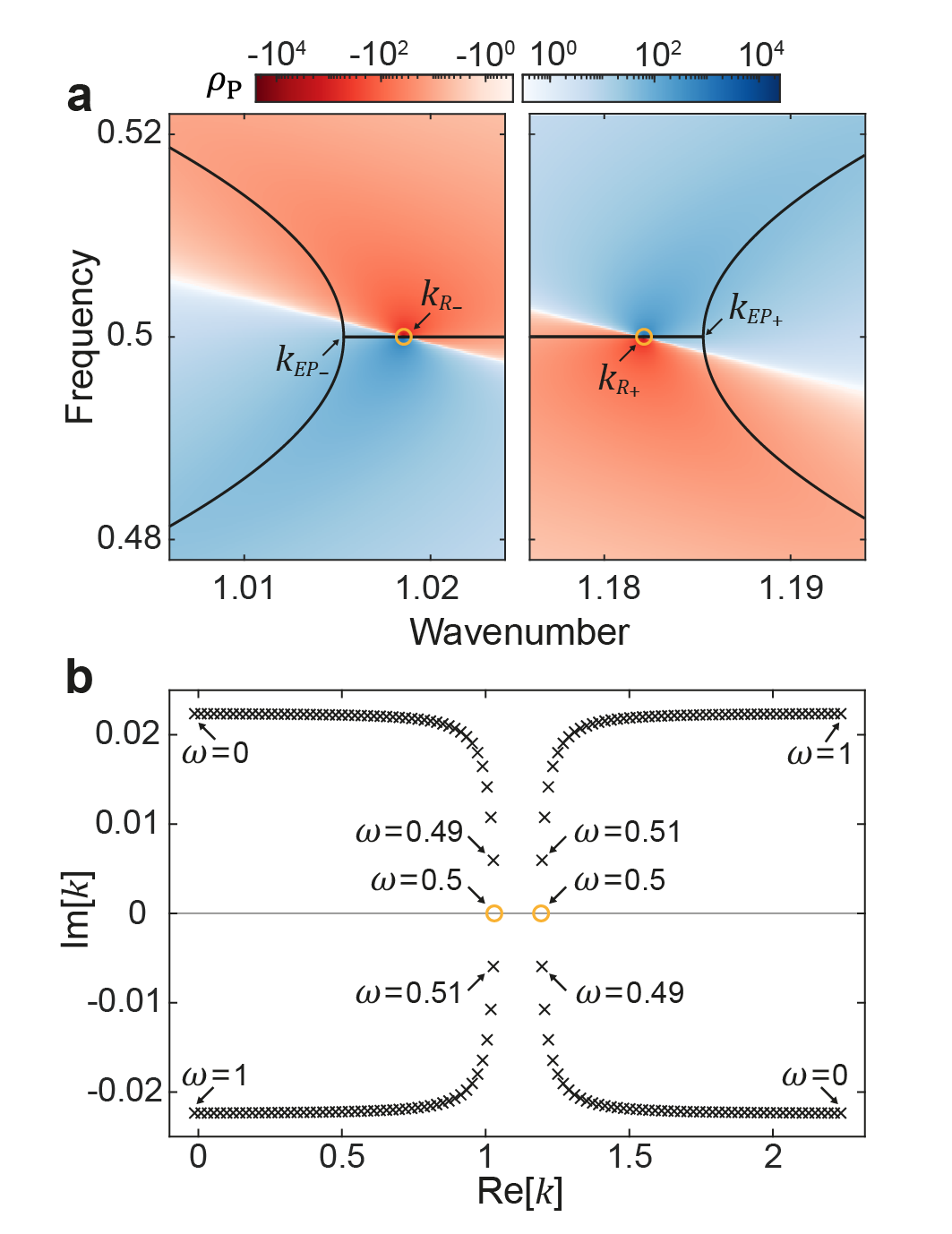}
  \caption{\label{fig:Fig3} (a) At the momentum gap frequency, the exceptional points are located at \(k_{EP_\pm}\), whereas the simple poles reside at \(k_{R_\pm}\). Blue (red) regions indicate positive (negative) $\mathrm{kDOS}$, corresponding to decay (excitation).  (b) Trajectories of the singularities in the complex-\(k\) plane as the emitter frequency is detuned. Each pole (\(\circ\)) leaves the real axis and becomes a branch point (\(\times\)), after which the rate integrals converge.}
\end{figure}

While the spontaneous emission decay and excitation rates follow from Eqs. \eqref{eqn:SED_rate} and \eqref{eqn:SEE_rate}, evaluating them at the gap frequency for the low loss regime is complicated by singularities in the full Green's function. Specifically, a pair of simple poles, $k_{R_-}$ and $k_{R_+}$ appear on the real wavenumber axis within the gap (Fig.~\ref{fig:Fig3}a), and at these points the full Green's function as well as the decay and excitation rates become unbounded. As the loss decreases, the simple pole moves toward the exceptional point, and in the lossless limit a simple pole and a double pole coexist at that point. When the emitter frequency is detuned slightly from the gap frequency, the simple poles leave the real \(k\) axis and become branch points in the complex \(k\) plane (Fig.~\ref{fig:Fig3}b). Once these singularities lie off the real axis, the full Green's function is analytic along the integration path and the spontaneous emission rate integral converges.


\begin{figure}[h!]
  \centering
    \includegraphics[width=0.45\textwidth]{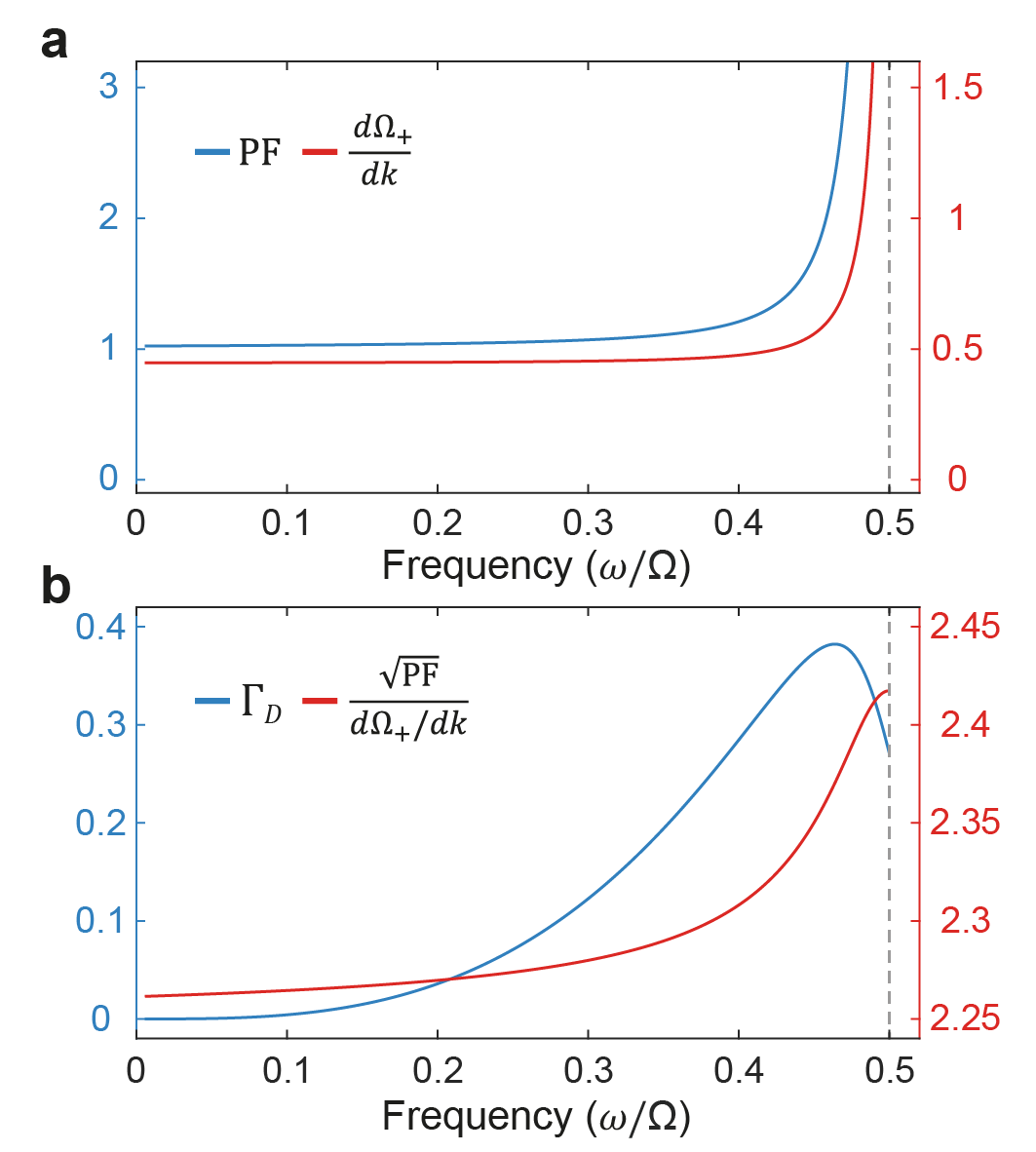}
  \caption{\label{fig:Fig5}(a) As \(\omega\) approaches the momentum gap frequency, the Petermann factor \(\mathrm{PF}\) and the dispersion slope \(d\Omega_{+}/dk\) both diverge; their asymptotic trends are shown. (b) With the singular contributions removed, the spontaneous emission decay rate converges to a finite value at the gap frequency.  The ratio \(\mathrm{PF}\!\big/\!\bigl(d\Omega_{+}/dk\bigr)\) is plotted alongside the rate and likewise converges near the gap frequency. }
\end{figure}

Although the singularities make the rate formally divergent, the spontaneous emission decay rate in a lossless PTC can still be estimated by omitting their contribution and taking the asymptotic limit \(\sigma \to 0\).  Under this approximation, the $\mathrm{kDOS}$ outside the gap \( (\,|\mathbf{k}|<k_{EP_-})\) becomes 
\begin{equation}
    \lim_{\sigma\to0}\rho_\mathbf{p}(\mathbf{k},\omega)\approx\sum_{\alpha=\pm}\delta(\omega-\Omega_\alpha)\mathrm{Re}[I_\alpha (\mathbf{k})],
\end{equation}
from which the \textit{remnant} spontaneous emission decay rate is given by
\begin{equation}
\mathrm{\Gamma}_D(\omega)\approx\frac{\omega|\mathbf{\tilde{p}}|^2}{8\hbar\epsilon_0}|\mathbf{k}|^2 \mathrm{Re}[\mathcal{I}_+] \frac{\sqrt{\mathrm{PF}}}{d\Omega_+/d|\mathbf{k}|}.\label{eqn:DOS_slope_PF}
\end{equation}
Here, the Petermann factor is (see Supplemental Material~F \cite{SM})
\begin{equation}
    \mathrm{PF}=\frac{\braket{R_+(\mathbf{k})}{R_+(\mathbf{k})}\braket{L_+(\mathbf{k})}{L_+(\mathbf{k})}}{|\braket{L_+(\mathbf{k})}{R_+(\mathbf{k})}|^2},
\end{equation}
and \(\mathcal{I}_{+}\equiv I_{+}(\mathbf{k})/\sqrt{\mathrm{PF}}\) is evaluated for \(\mathbf{k}\perp\mathbf{n}_{\mathbf{p}}\). As \(\omega\) approaches \(\Omega/2\) from below, both \(\sqrt{\mathrm{PF}}\) and the dispersion slope \(d\Omega_{+}/d|\mathbf{k}|\) diverge, yet their ratio remains finite.  Equivalently, \(\mathrm{Re}[I_{+}(\mathbf{k})]\big/\!\bigl(d\Omega_{+}/d|\mathbf{k}|\bigr)\) stays bounded (see Supplemental Material G for the $\mathrm{kDOS}$ at the exceptional points \cite{SM}). Hence, the spontaneous emission decay rate converges to a finite positive value.  This behavior contradicts the claim that the decay vanishes at the band edge because the dispersion slope is vertical \cite{doi:10.1126/science.abo3324}.  The Petermann factor counterbalances the slope divergence and underscores the influence of non-orthogonal photonic Floquet modes. Consistent with our analysis, a non-zero decay rate has already been confirmed, both theoretically and experimentally, in time-invariant photonic systems that exhibit similarly steep band edges \cite{Pick2017,PhysRevLett.129.083602}.

The study of spontaneous emission in time-periodic photonic structures has deepened our understanding of non-equilibrium light-matter interactions. We show that tailoring the temporal environment enhances the spontaneous emission decay rate, particularly at the momentum gap frequency of PTCs, adding a temporal dimension to emission control beyond Hermitian regime \cite{doi:10.1126/science.abo3324,https://doi.org/10.1002/adom.202402444,PhysRevB.111.125419}. This idea provides a new control knob for light-matter interaction, going beyond conventional approaches of past decades, which have relied solely on resonant or spatially periodic structures surrounding atoms or quantum emitters \cite{1972JETP...35..269B,Bykov1975,PhysRevLett.58.2059,JOANNOPOULOS1997165,PhysRevLett.78.3294,PhysRevLett.81.77,Boroditsky:99,PhysRevLett.84.4341,Lodahl2004,PhysRevLett.95.013904,altug2006ultrafast,PhysRevLett.99.023902,Noda2007,PhysRevLett.99.193901,novotny2012principles,PhysRevLett.110.237401,PhysRevLett.117.107402,barnes2020classical}. Time periodicity introduces loss, gain, and Floquet eigenmode non-orthogonality, revealing a richer physics than previously envisioned. Notably, the newly identified spontaneous excitation with photon emission challenges conventional theories, distinguishing equilibrium and non-equilibrium emission processes. These findings highlights the need for a more detailed quantum electrodynamic framework to fully describe PTCs.

\begin{acknowledgments}
We would like to thank Prof. Hansuek Lee for the helpful discussions. This work was supported by National Research Foundation of Korea (NRF) through the government of Korea (NRF- 2022R1A2C301335313) and the Samsung Science and Technology Foundation (SSTF-BA2402-02). H.-C.P. acknowledges the support by NRF (No. RS-2023-00278511). J.-W.R. acknowledges financial support from the Institute for Basic Science in the Republic of Korea (Project IBS-R024-D1). G.Y.C. is supported by the NRF (Grant No.RS-2023-00208291) funded by the Korean Government (MSIT). N.P. acknowledges financial support from the NRF through the Midcareer Researcher Program (No. RS-2023-00274348). J.S. acknowledges the support by NRF (2021R1A2C2008687). The work done in Hong Kong is supported by the RGC of Hong Kong through the grant AoE/P-502/20.
\end{acknowledgments}

\bibliography{refs}

\clearpage 
\begin{widetext}

\setcounter{equation}{0} 
\setcounter{figure}{0} 
\renewcommand{\theequation}{S\arabic{equation}} 
\renewcommand{\thefigure}{S\arabic{figure}} 

{\centering
\large \textbf{Supplemental Material: Spontaneous emission decay and excitation \\in photonic time crystals} \\[2ex]  

\normalsize Jagang Park$^{1, \ast}$, Kyungmin Lee$^{2, \ast}$, Ruo-Yang Zhang$^{3}$, Hee-Chul Park$^{4}$, Jung-Wan Ryu$^{4}$, Gil Young Cho$^{5}$, Min Yeul Lee$^{6}$, Zhaoqing Zhang$^{3}$, Namkyoo Park$^{7}$, Wonju Jeon$^{8}$, Jonghwa Shin$^{6}$, C. T. Chan$^{3}$, and Bumki Min$^{2,\dagger}$ \\[1ex] 
\fontsize{9pt}{10pt}\selectfont $^{1}$\textit{Department of Electrical Engineering and Computer Sciences,\\ University of California, Berkeley, California 94720, USA} \\
$^{2}$\textit{Department of Physics, Korea Advanced Institute of Science and Technology, Daejeon 34141, Republic of Korea} \\
$^{3}$\textit{Department of Physics, the Hong Kong University of Science and Technology,\\ Clear Water Bay, Kowloon, Hong Kong 999077, China} \\
$^{4}$\textit{Center for Theoretical Physics of Complex Systems,\\ Institute for Basic Science, Daejeon 34126, Republic of Korea} \\
$^{5}$\textit{Department of Physics, Pohang University of Science and Technology, Pohang 37673, Republic of Korea} \\
$^{6}$\textit{Department of Material Sciences and Engineering,\\ Korea Advanced Institute of Science and Technology, Daejeon 34141, Republic of Korea} \\
$^{7}$\textit{Department of Electrical and Computer Engineering,\\ Seoul National University, Seoul 08826, Republic of Korea} \\
$^{8}$\textit{Department of Mechanical Engineering, Korea Advanced Institute of Science and Technology,\\ 
Daejeon 34141, Republic of Korea} \\
\textit{$^{\ast}$These authors contributed equally to this work. $^{\dagger}$bmin@kaist.ac.kr} \\
}

\renewcommand{\thesection}{\Alph{section}}
\counterwithin{figure}{section}

\section{Dyadic Green's function for PTCs}
In this section, we derive the dyadic Green's function for PTCs in the wavenumber-frequency space. Maxwell's equations are reformulated into a Schrödinger-like equation with a source term as follows:
\begin{equation}
    i\frac{\partial}{\partial t}\begin{bmatrix} \mathbf{E}(\mathbf{r},t)\\ \mathbf{H}(\mathbf{r},t)\end{bmatrix}=\begin{bmatrix} -i\left(\frac{\dot{\epsilon}(t)+\sigma}{\epsilon(t)}\right)\mathbf{I}_3 & \frac{i}{\epsilon(t)}\nabla\times\\-\frac{i}{\mu}\nabla\times & \mathbf{0}_3\end{bmatrix}\begin{bmatrix} \mathbf{E}(\mathbf{r},t)\\ \mathbf{H}(\mathbf{r},t)\end{bmatrix}-\frac{i}{\epsilon(t)}\begin{bmatrix} 1\\0 \end{bmatrix}\otimes \mathbf{J}(\mathbf{r},t) \label{eqn:1st-order form}
\end{equation}
where $\mathbf{I}_3$ and $\mathbf{O}_3$ denote the $3 \times 3$ identity matrix and the null matrix, respectively. The spatial homogeneity of PTCs renders the formulation of the Schrödinger-like equation in momentum space particularly advantageous. The spatial Fourier transformation of Eq. \eqref{eqn:1st-order form} results in
\begin{equation}
    i\frac{\partial}{\partial t}\ket{\Psi(\mathbf{k},t)}=\mathcal{H}(\mathbf{k},t)\ket{\Psi(\mathbf{k},t)}-iA(t)\otimes \mathbf{J}(\mathbf{k},t), \label{eqn:1st-order form in k space}
\end{equation}
where 
\begin{equation}
    \ket{\Psi(\mathbf{k},t)}=\begin{bmatrix} \mathbf{E}(\mathbf{k},t)\\ \mathbf{H}(\mathbf{k},t)\end{bmatrix},\, \mathcal{H}(\mathbf{k},t)= \begin{bmatrix} -i\left(\frac{\dot{\epsilon}(t)+\sigma}{\epsilon(t)}\right)\mathbf{I}_3 & -\frac{1}{\epsilon(t)}\mathbf{k}\times\\\frac{1}{\mu}\mathbf{k}\times & \mathbf{0}_3\end{bmatrix}, \, A(t) = \frac{1}{\epsilon(t)}\begin{bmatrix} 1\\0 \end{bmatrix}.
\end{equation}
Fourier transforming the above equation in the time domain yields 
\begin{equation}
    \omega\ket{\Psi(\mathbf{k},\omega)}=\frac{1}{2\pi}\int d\omega'\mathcal{H}(\mathbf{k},\omega-\omega')\ket{\Psi(\mathbf{k},\omega')}-\frac{i}{2\pi}\int d\omega'A(\omega-\omega')\otimes\mathbf{J}(\mathbf{k},\omega').
\end{equation}
We can now define the dyadic Green's function $\mathbb{G}(\mathbf{k},\omega,\omega_0)\equiv[\mathbb{G}^E ; \mathbb{G}^H]$, which satisfies the following equation,
\begin{equation}
    \omega\mathbb{G}(\mathbf{k},\omega,\omega_0)=\frac{1}{2\pi}\int d\omega'\mathcal{H}(\mathbf{k},\omega-\omega')\mathbb{G}(\mathbf{k},\omega',\omega_0)-\frac{1}{2\pi}\int d\omega'(\mu\omega')^{-1}A(\omega-\omega')\otimes\mathbb{I}\delta(\omega'-\omega_0)\label{eqn:dyadic greens function}
\end{equation}
where $\mathbb{I}$ is the unit dyad. Therefore, the general solution for the field vector $\ket{\Psi(\mathbf{k},\omega)}$, associated with an arbitrary current source $\mathbf{J}(\mathbf{k},\omega')$, is given by $\ket{\Psi(\mathbf{k},\omega)}=\int d\omega'\mathbb{G}(\mathbf{k},\omega,\omega')i\mu\omega'\mathbf{J}(\mathbf{k},\omega')$. Here, due to the temporal periodicity of the permittivity, $\epsilon(t)=\epsilon(t+2\pi/\Omega)$, we can express $\mathcal{H}(\mathbf{k},\omega)$ and $A(\omega)$ as $\sum_m\mathcal{H}(\mathbf{k},\omega)\delta(\omega/\Omega-m)$ and $\sum_m A(\omega)\delta(\omega/\Omega-m)$, respectively. This leads to a simplification of Eq. \eqref{eqn:dyadic greens function} to
\begin{equation}
    \omega\mathbb{G}(\mathbf{k},\omega,\omega_0)=\frac{1}{2\pi}\sum_m\mathcal{H}(\mathbf{k},m\Omega)\mathbb{G}(\mathbf{k},\omega-m\Omega,\omega_0)-\frac{1}{2\pi\mu\omega_0}A(\omega-\omega_0)\otimes\mathbb{I}.
\end{equation}
Additionally, by incorporating the temporal periodicity into the Green's function, $\mathbb{G}(\mathbf{k},\omega,\omega')$ can be represented as $\sum_m\mathbb{G}(\mathbf{k},\omega,\omega')\delta(\omega/\Omega-\omega'/\Omega-m)$. Integrating over $\omega$ then results in
\begin{equation}
    (\omega_0+n\Omega)\mathbb{G}_n(\mathbf{k},\omega_0)=\sum_m\mathcal{H}_m(\mathbf{k})\mathbb{G}_{n-m}(\mathbf{k},\omega_0)-\frac{1}{\mu\omega_0}A_n\otimes\mathbb{I},
\end{equation}
where $\mathbb{G}_m(\mathbf{k},\omega_0)$ is defined to be $\mathbb{G}(\mathbf{k},\omega_0+m\Omega,\omega_0)$. Both $\mathcal{H}(\mathbf{k},t)$ and $A(t)$ can be represented as Fourier series, with $\mathcal{H}(\mathbf{k},t) = \sum_m \mathcal{H}_m(\mathbf{k}) e^{-im\Omega t}$ and $A(t) = \sum_m A_m e^{-im\Omega t}$, where the coefficients are given by $\mathcal{H}_m(\mathbf{k}) = \frac{1}{2\pi} \mathcal{H}(\mathbf{k},m\Omega)$ and $A_m = \frac{1}{2\pi} A(m\Omega)$, respectively. This allows us to formulate the system of linear equations in matrix form as follows, 
\begin{equation}
\omega_0\mathbb{G}_F(\mathbf{k},\omega_0)=\mathcal{H}_F(\mathbf{k})\mathbb{G}_F(\mathbf{k},\omega_0)-\frac{1}{\mu\omega_0}A_F\otimes\mathbb{I},\label{eqn:dyadic greens function matrix form}
\end{equation}
where 
\begin{equation}
    \mathbb{G}_F(\mathbf{k},\omega_0) = \begin{bmatrix}
        \vdots \\
        \mathbb{G}_{-1}(\mathbf{k},\omega_0)\\
        \mathbb{G}_{0}(\mathbf{k},\omega_0)\\
        \mathbb{G}_{+1}(\mathbf{k},\omega_0)\\
        \vdots
    \end{bmatrix}\text{,}\,
    \mathcal{H}_F(\mathbf{k}) = \begin{bmatrix}
        \ddots & \ddots & & & \\
        \ddots & \mathcal{H}_0(\mathbf{k})+\Omega \mathbf{I}_6 & \mathcal{H}_{-1}(\mathbf{k}) & \mathcal{H}_{-2}(\mathbf{k}) & \\
               & \mathcal{H}_{+1}(\mathbf{k}) & \mathcal{H}_0(\mathbf{k}) & \mathcal{H}_{-1}(\mathbf{k}) & \\
               &    \mathcal{H}_{+2}(\mathbf{k})         & \mathcal{H}_{+1}(\mathbf{k}) & \mathcal{H}_0(\mathbf{k})-\Omega \mathbf{I}_6 & \ddots \\
               &    &   & \ddots & \ddots
    \end{bmatrix} \label{eqn:Floquet_Ham_3D} 
\end{equation}
and $A_F = [\cdots;A_{-1};A_0;A_{+1};\cdots ]$. As a result, $\mathbb{G}_F(\mathbf{k},\omega_0)$ can be expressed as
\begin{equation}
\begin{aligned}
    \mathbb{G}_F(\mathbf{k},\omega_0) &=-\frac{1}{\mu\omega_0}[\omega_0\mathbf{I}_F-\mathcal{H}_F(\mathbf{k})]^{-1}\cdot A_F\otimes\mathbb{I}\\
    &\equiv -\frac{1}{\mu\omega_0}\mathcal{G}_F(\mathbf{k},\omega_0)\cdot A_F\otimes\mathbb{I},
\end{aligned}
\end{equation}
where $\mathbf{I}_F$ denotes the identity matrix matching the dimensions of $\mathcal{H}_F(\mathbf{k})$, and $\mathcal{G}_F(\mathbf{k},\omega_0)$ represents the Green's function within the Floquet formalism. One can determine $\mathbb{G}^E_n(\mathbf{k},\omega_0)$ $(\equiv\mathbb{G}^E(\mathbf{k},\omega_0+n\Omega,\omega_0))$ by selecting the corresponding elements from $\mathcal{G}_F(\mathbf{k},\omega_0)$, i.e.,
\begin{equation}
    \mathbb{G}^E_n(\mathbf{k},\omega_0)=-\frac{1}{\mu\omega_0}\delta_n\cdot\mathcal{G}_F(\mathbf{k},\omega_0)\cdot A_F\otimes\mathbb{I}
\end{equation}
where $\delta_n=[\cdots,\delta_{n,m-1},\delta_{n,m},\delta_{n,m+1},\cdots]\otimes[1,0]\otimes\mathbb{I}$, and $\delta_{n,m}$ denotes the Kronecker delta. In the main text, we omitted the superscript $E$ from the notation of the dyadic Green's function for simplicity.

\section{Derivation of $\mathrm{kDOS}$}
The oscillating point electric dipole, $\mathbf{p}(\mathbf{r},t) = \mathrm{Re}[\mathbf{p}\delta(\mathbf{r}-\mathbf{r}_0)e^{-i\omega_0t}]$, can be decomposed into a superposition of extended dipoles. This decomposition is given by $\mathbf{p}(\mathbf{r},t) = \int \mathbf{p}_{\mathbf{k}_0}(\mathbf{r},t) \, d^3\mathbf{k}_0$, where the extended dipole at each wavevector is defined as $\mathbf{p}_{\mathbf{k}_0}(\mathbf{r},t) = (2\pi)^{-3}\mathrm{Re}[\mathbf{p}e^{i(\mathbf{k}_0\cdot(\mathbf{r}-\mathbf{r}_0)-\omega_0t)}]$. The spatial homogeneity of the PTCs ensures that electromagnetic fields associated with different wavevectors $\mathbf{k}$ are independent. Consequently, the time-averaged power dissipated by each extended dipole, $\bar{P}(\mathbf{k}_0,\omega_0)$, can be calculated individually. Integrating $\bar{P}(\mathbf{k}_0,\omega_0)$ over the entire wavevector space results in the total time-averaged dissipated power, $\bar{P}(\omega_0) = \int_{\mathbb{R}^3} \bar{P}(\mathbf{k}_0,\omega_0) \, d^3\mathbf{k}_0$. We now turn our attention to the current density induced by the extended dipole, which is defined as $\mathbf{J}_{\mathbf{k}_0}(\mathbf{r},t) = \partial \mathbf{p}_{\mathbf{k}_0}(\mathbf{r},t)/\partial t$. After performing a Fourier transform, this current density can be expressed as:
\begin{equation}
\mathbf{J}_{\mathbf{k}_0}(\mathbf{k},\omega) = -i\omega_0\pi \left[ p e^{-i\mathbf{k}_0\cdot\mathbf{r}_0} \delta(\omega-\omega_0)\delta(\mathbf{k}-\mathbf{k}_0) - p^* e^{i\mathbf{k}_0\cdot\mathbf{r}_0} \delta(\omega+\omega_0)\delta(\mathbf{k}+\mathbf{k}_0) \right] \mathbf{n}_{\mathbf{p}},\label{eqn:extended source in k space}
\end{equation}
where $p$ represents the complex amplitude of the dipole, and $\mathbf{n}_{\mathbf{p}}$ is the unit vector indicating the direction of the dipole moment. The electric field resulting from the extended current density can be calculated using the dyadic Green's function as follows:
\begin{equation}
\begin{aligned}
    \mathbf{E}(\mathbf{k},\omega) &= \int d\omega' \mathbb{G}^E(\mathbf{k},\omega,\omega')i\mu\omega'\mathbf{J}_{\mathbf{k}_0}(\mathbf{k},\omega')\\
    &=\mu\pi\omega_0^2\{p e^{-i\mathbf{k}_0\cdot\mathbf{r}_0}\mathbb{G}^E(\mathbf{k},\omega,\omega_0)\delta(\mathbf{k}-\mathbf{k}_0)+p^* e^{i\mathbf{k}_0\cdot\mathbf{r}_0}\mathbb{G}^E(\mathbf{k},\omega,-\omega_0)\delta(\mathbf{k}+\mathbf{k}_0)\} \mathbf{n}_{\mathbf{p}}.
\end{aligned}\label{eqn:E field  by an extended source in k space}
\end{equation}
The power dissipated by an extended dipole can be calculated as follows:
\begin{equation}
\begin{aligned}
    P(t)&=-\int d^3\mathbf{r} \mathbf{J}_{\mathbf{k}_0}(\mathbf{r},t)\cdot\mathbf{E}(\mathbf{r},t)\\
    &=-\int d^3\mathbf{r}\frac{1}{(2\pi)^8}\int d^3 \mathbf{k} d\omega e^{i(\mathbf{k}\cdot\mathbf{r}-\omega t)}\mathbf{J}_{\mathbf{k}_0}(\mathbf{k},\omega)*\mathbf{E}(\mathbf{k},\omega)\\
    &=-\int d^3\mathbf{r}\frac{1}{(2\pi)^8}\int d^3 \mathbf{k} d\omega e^{i(\mathbf{k}\cdot\mathbf{r}-\omega t)}\int d^3\mathbf{k'} d\omega'\mathbf{J}_{\mathbf{k}_0}(\mathbf{k}-\mathbf{k'},\omega-\omega')\cdot\mathbf{E}(\mathbf{k'},\omega')\\
    &=-\frac{1}{(2\pi)^5}\int d^3 \mathbf{k} d\omega \delta(\mathbf{k}) e^{-i\omega t}\int d^3\mathbf{k'} d\omega'\mathbf{J}_{\mathbf{k}_0}(\mathbf{k}-\mathbf{k'},\omega-\omega')\cdot\mathbf{E}(\mathbf{k'},\omega')\\
    &=-\frac{1}{(2\pi)^5}\int d\omega e^{-i\omega t}\int d^3\mathbf{k'} d\omega'\mathbf{J}_{\mathbf{k}_0}(-\mathbf{k'},\omega-\omega')\cdot\mathbf{E}(\mathbf{k'},\omega').
    \end{aligned}\label{eqn:radiated power from extended dipole}
\end{equation}
By incorporating Eq. \eqref{eqn:extended source in k space} and Eq. \eqref{eqn:E field by an extended source in k space} into Eq. \eqref{eqn:radiated power from extended dipole}, the dissipated power can be determined as follows:
\begin{equation}
    P(t)=-\frac{1}{(2\pi)^5}\int d\omega e^{-i\omega t}i\pi^2\omega_0^3\mu|\mathbf{p}|^2\{\mathbf{n}_\mathbf{p}\cdot\mathbb{G}^E(\mathbf{k}_0,\omega+\omega_0,\omega_0)\cdot\mathbf{n}_\mathbf{p}-\mathbf{n}_\mathbf{p}\cdot\mathbb{G}^E(-\mathbf{k}_0,\omega-\omega_0,-\omega_0)\cdot\mathbf{n}_\mathbf{p}\}.\label{eqn:radiated power from extended dipole2}
\end{equation}
By noting that $\mathbb{G}(\mathbf{k},\omega,\omega')=\sum_m\mathbb{G}(\mathbf{k},\omega,\omega')\delta(\omega/\Omega-\omega'/\Omega-m)$ and using the property $\mathbb{G}(\mathbf{k},\omega,\omega')=\mathbb{G}^*(-\mathbf{k},-\omega,-\omega')$, the dissipated power can be expressed as:
\begin{equation}
    P(t)=\frac{1}{(2\pi)^3}\frac{\omega_0^3\mu|\mathbf{p}|^2}{2}\sum_n\mathbf{n}_\mathbf{p}\cdot\mathrm{Im}[e^{-in\Omega t}\mathbb{G}^E_n(\mathbf{k_0},\omega_0)]\cdot\mathbf{n}_\mathbf{p}.\label{eqn:radiated power from extended dipole3}
\end{equation}
Upon time-averaging, only the term with $n=0$ remains significant. Consequently, the time-averaged power dissipated by an extended dipole, characterized by a wavevector $\mathbf{k}_0$ and frequency $\omega_0$, can be expressed as
\begin{equation}
\begin{aligned}
    \bar{P}(\mathbf{k}_0,\omega_0)&=\frac{1}{(2\pi)^3}\frac{\omega_0^3\mu|\mathbf{p}|^2}{2}\mathbf{n}_\mathbf{p}\cdot\mathrm{Im}[\mathbb{G}^E_0(\mathbf{k_0},\omega_0)]\cdot\mathbf{n}_\mathbf{p}\\
    &\equiv \frac{1}{(2\pi)^3}\frac{\pi\omega^2|\mathbf{p}|^2}{4\epsilon_0}\rho_\mathbf{p}(\mathbf{k}_0,\omega_0).
\end{aligned}
\end{equation}
Here, we define $\rho_\mathbf{p}(\mathbf{k},\omega)$ as the partial momentum-resolved photonic density of states ($\mathrm{kDOS}$), which is expressed as 
\begin{equation}
\rho_\mathbf{p}(\mathbf{k},\omega)= \frac{2\epsilon_0\mu\omega}{\pi }\mathrm{Im}[\mathbf{n}_\mathbf{p}\cdot\mathbb{G}_0^E(\mathbf{k},\omega)\cdot\mathbf{n}_\mathbf{p}].
\end{equation}
Figure \ref{fig:sup_kDOS} presents the $\mathrm{kDOS}$ map and the nonradiative power dissipation for a PTC characterized by a sinusoidally-modulated relative permittivity $\epsilon(t) = \epsilon_0 + \epsilon_m \sin(\Omega t)$, where $\epsilon_0 = 5$, $\epsilon_m = 1.5$, $\mu = 1$, $\Omega = 1$, and $\sigma = 0.1$ (less than the critical value $\sigma_c$, indicating a small loss case). In Fig. \ref{fig:sup_kDOS}a, the $\mathrm{kDOS}$ is plotted for the case where the wavevector $\mathbf{k}$ is perpendicular to the dipole orientation vector $\mathbf{n}_\mathbf{p}$. Additionally, Fig. \ref{fig:sup_kDOS}b shows the nonradiative power dissipation map calculated when $\mathbf{k}$ is parallel to $\mathbf{n}_\mathbf{p}$. Notably, the nonradiative power dissipation for the parallel alignment case diminishes to zero as the conductivity approaches zero ($\sigma \rightarrow 0$).
\clearpage
\begin{figure}[htb!]
  \centering
    \includegraphics[width=0.9\textwidth]{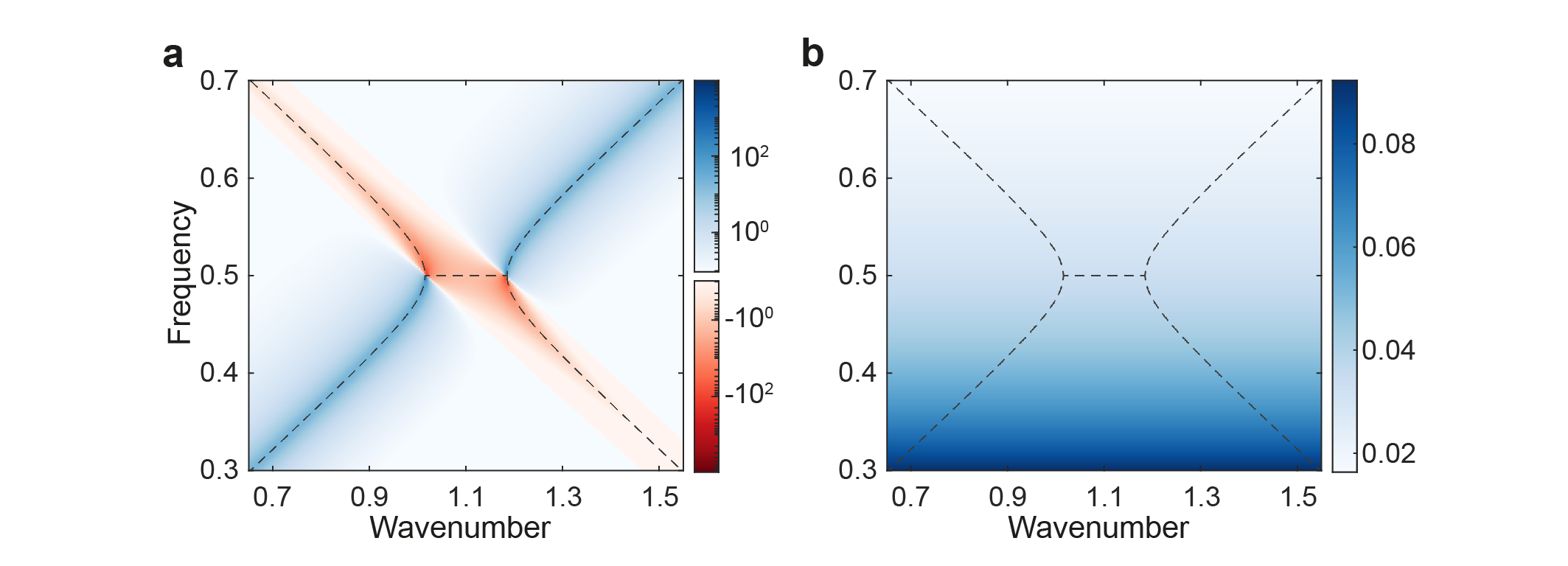}
  \caption{\label{fig:sup_kDOS}$\mathrm{kDOS}$ and nonradiative power dissipation maps for a PTC with sinusoidally-modulated relative permittivity, $\epsilon(t) = \epsilon_0 + \epsilon_m \sin(\Omega t)$, where the parameters are set to $\epsilon_0 = 5$, $\epsilon_m = 1.5$, $\mu = 1$, $\Omega = 1$, and $\sigma = 0.1$, indicating a small loss scenario ($\sigma < \sigma_c$). Panel (a) shows the $\mathrm{kDOS}$ when the wavevector $\mathbf{k}$ is perpendicular to the dipole orientation vector $\mathbf{n}_\mathbf{p}$. Panel (b) shows the nonradiative power dissipation when $\mathbf{k}$ is parallel to $\mathbf{n}_\mathbf{p}$.}
\end{figure}

\section{Pseudo-Hermiticity of PTCs}
Under the source-free condition, Maxwell's equations for a plane electromagnetic wave can be reformulated into the following matrix differential equation:
\begin{equation}
    \begin{bmatrix} 0 & k\\k & 0\end{bmatrix}\begin{bmatrix} E\\H\end{bmatrix}=-i\frac{\partial}{\partial t}\begin{bmatrix} \epsilon(t) & 0\\0 & \mu\end{bmatrix}\begin{bmatrix} E\\H\end{bmatrix}-i\begin{bmatrix} \sigma & 0\\0 & 0\end{bmatrix}\begin{bmatrix} E\\H\end{bmatrix}.\label{eqn:Pseudo-Hermitian1}
\end{equation}
Due to the time-periodic variation in permittivity, the solutions to Eq. \eqref{eqn:Pseudo-Hermitian1}  can be represented as a Floquet mode: $\ket{\Psi(t)} = [E \; H]^T = e^{-i\omega t}\ket{\Phi(t)}$, where $\ket{\Phi(t)}=\ket{\Phi(t+T)}$. Substituting $\ket{\Psi(t)}$ into Eq. \eqref{eqn:Pseudo-Hermitian1} yields
\begin{equation}
    \omega\ket{\Phi(t)}=\left[-i\frac{\partial}{\partial t}-\frac{k}{\mu\epsilon (t)}\begin{bmatrix} 0 & \mu\\ \epsilon (t) & 0\end{bmatrix}-\frac{i\{\dot{\epsilon} (t)+\sigma\}}{2\epsilon (t)}(\mathbb{I}_2+\mathbf{\tau}_z) \right]\ket{\Phi (t)},\label{eqn:Pseudo-Hermitian2}
\end{equation}
where $\mathbb{I}_2$ denotes the identity matrix and $\mathbf{\tau}_i$ represents the Pauli matrix. Next we define a non-periodic mode $\ket{\Phi_N (t)}$ as follows:
\begin{equation}
\begin{aligned}
    \ket{\Phi_N (t)}&\equiv\exp(\int\frac{\dot{\epsilon}(t)+\sigma}{2\epsilon(t)}dt)\ket{\Phi (t)}\\
    &=\exp(\frac{a_0\sigma}{2}t)\exp{\frac{1}{2}\ln{\epsilon(t)}+i\frac{\sigma}{2}\sum_{n\neq0}\frac{a_n}{n\Omega}e^{-in\Omega}}\ket{\Phi (t)}\\
    &\equiv\exp(\frac{a_0\sigma}{2}t)\ket{\Phi_P (t)},\label{eqn:Pseudo-Hermitian3}
\end{aligned}
\end{equation}
where the $a_n$'s are the Fourier expansion coefficients of $1/\epsilon(t)$, i.e., $1/\epsilon(t)=\sum_na_ne^{-in\Omega t}$. Here, it can be shown that $\ket{\Phi_P (t)}$ is time-periodic with the periodicity of $T$, i.e., $\ket{\Phi_P (t)}=\ket{\Phi_P (t+T)}$. Then, Eqs. \eqref{eqn:Pseudo-Hermitian2} and \eqref{eqn:Pseudo-Hermitian3} can be reformulated into the following eigenvalue equation:
\begin{equation}
\begin{aligned}
    \left(\omega+\frac{ia_0\sigma}{2}\right)\ket{\Phi_P (t)}&=\left[-i\frac{\partial}{\partial t}-\frac{k}{\mu\epsilon (t)}\begin{bmatrix} 0 & \mu\\ \epsilon (t) & 0\end{bmatrix}-\frac{i\{\dot{\epsilon} (t)+\sigma\}}{2\epsilon (t)}\mathbf{\tau}_z \right]\ket{\Phi_P (t)}\\
    &=\mathcal{H}_{\mathrm{e}}(t)\ket{\Phi_P (t)},\label{eqn:Pseudo-Hermitian4}
\end{aligned}
\end{equation}
where the effective Hamiltonian matrix $\mathcal{H}_{\mathrm{e}}(t)$ can be shown to be $\mathbf{\tau}_x$-pseudo-Hermitian, i.e., $\mathcal{H}^\dagger_{\mathrm{e}}=\mathbf{\tau}_x\mathcal{H}_{\mathrm{e}}\mathbf{\tau}_x^{-1}$.

\section{Analysis with driven Lorentz model}
In this section, we utilize a driven Lorentz oscillator model to demonstrate that we can obtain results qualitatively similar to those presented in the main text. Within this model, the polarization density $\mathbf{P}$ and the electric field $\mathbf{E}$ are described by the following equation:
\begin{equation}
\frac{\partial^2\mathbf{P}}{\partial t^2} + \eta\frac{\partial\mathbf{P}}{\partial t} + \frac{\kappa(t)}{m}\mathbf{P} = \frac{Ne^2}{m}\mathbf{E}, \label{eqn:LZmodel}
\end{equation}
where $N$ represents the number of atoms per unit volume, $m$ denotes the mass of the bound charge, $\eta$ is the damping coefficient, and $e$ is the elementary charge. The elastic constant, $\kappa(t)$, is considered to be time-periodic with a period $T = 2\pi/\Omega$. The PTC can be analyzed by solving Maxwell's equations in conjunction with the driven Lorentz model:
\begin{equation}
\begin{aligned} 
    \nabla\times \mathbf{H}&=\frac{\partial\mathbf{D}}{\partial t}+\mathbf{J}, \quad\mathbf{D}=\epsilon_{v}\mathbf{E}+\mathbf{P},\\
    \nabla\times \mathbf{E}&=-\frac{\partial\mathbf{B}}{\partial t},\quad\mathbf{B} = \mu_{v}\mathbf{H}.\label{eqn:Maxwell_lorentzian}  
\end{aligned} 
\end{equation}
To simplify the analysis, we focus here on the one-dimensional case. Considering an extended current source $\mathbf{J}(\mathbf{r},t) = \mathrm{Re}[Je^{i(kx-\omega t)}]\hat{\mathbf{z}}$, we can assume the electric and magnetic fields, as well as the polarization density, as follows: $\mathbf{E}(\mathbf{r},t) = \mathrm{Re}[E(t)e^{ikx}]\hat{\mathbf{z}}$, $\mathbf{H}(\mathbf{r},t) = \mathrm{Re}[H(t)e^{ikx}]\hat{\mathbf{y}}$, and $\mathbf{P}(\mathbf{r},t) = \mathrm{Re}[P(t)e^{ikx}]\hat{\mathbf{z}}$. These assumptions simplify Eqs \eqref{eqn:LZmodel} and \eqref{eqn:Maxwell_lorentzian}  into the following matrix differential equation:
\begin{equation}
    i\frac{\partial}{\partial t}\ket{a(t)}=A(t)\ket{a(t)}+\ket{S(t)},
\end{equation}
where

\begin{equation}
    A(t) = \begin{bmatrix}
        0 & -\frac{k}{\epsilon_{v}} & 0 & -i\frac{1}{\epsilon_{v}}\\ 
        -\frac{k}{\mu_{v}} & 0 & 0 &0 \\ 
        0 & 0 & 0 & i \\
        i\frac{Ne^2}{m} & 0 & -i\frac{\kappa(t)}{m} & -i\eta
    \end{bmatrix},\ \ket{a(t)} = \begin{bmatrix}
        E(t) \\
        H(t) \\
        P(t) \\
        \dot{P}(t) 
    \end{bmatrix},
\end{equation}
and

\begin{equation}
    \ket{S(t)} = -i\frac{J}{\epsilon_{v}}\begin{bmatrix}
        1\\0\\0\\0
    \end{bmatrix}e^{-i\omega t}.
\end{equation}
By defining the following matrix $R$ as,
\begin{equation}
    R = \begin{bmatrix}
        \sqrt{\epsilon_{v}} & \sqrt{\mu_{v}} & 0 & 0 \\ 
        \sqrt{\epsilon_{v}} & -\sqrt{\mu_{v}} & 0 & 0 \\
        0 & 0 & \sqrt{\frac{\kappa_0}{Ne^2}} & i\sqrt{\frac{m}{Ne^2}} \\
        0 & 0 & \sqrt{\frac{\kappa_0}{Ne^2}} & -i\sqrt{\frac{m}{Ne^2}}\\
    \end{bmatrix},
\end{equation}
Maxwell's equations can be transformed into a Schrödinger-like equation:
\begin{equation}
i\frac{\partial}{\partial t}
\ket{\Psi(t)}=\mathcal{H}(t)\ket{\Psi(t)}+\ket{s(t)},
\end{equation}
where $\mathcal{H}(t) = RA(t)R^{-1}$, $\ket{\Psi(t)} = R\ket{a(t)}$, and $\ket{s(t)}=R\ket{S(t)}=\ket{s}e^{-i\omega t}$. When the elastic constant remains constant (i.e., $\kappa(t) = \kappa_0$ in Eq. \eqref{eqn:LZmodel}), the band structure of the time-invariant medium can be determined using Eqs. \eqref{eqn:LZmodel} and \eqref{eqn:Maxwell_lorentzian}. The resulting band structure, shown in the top panel of Fig. \ref{fig:sup_LZ_band}, illustrates the presence of an energy gap resulting from the avoided crossing.

\begin{figure}[htb!]
  \centering
    \includegraphics[width=0.45\textwidth]{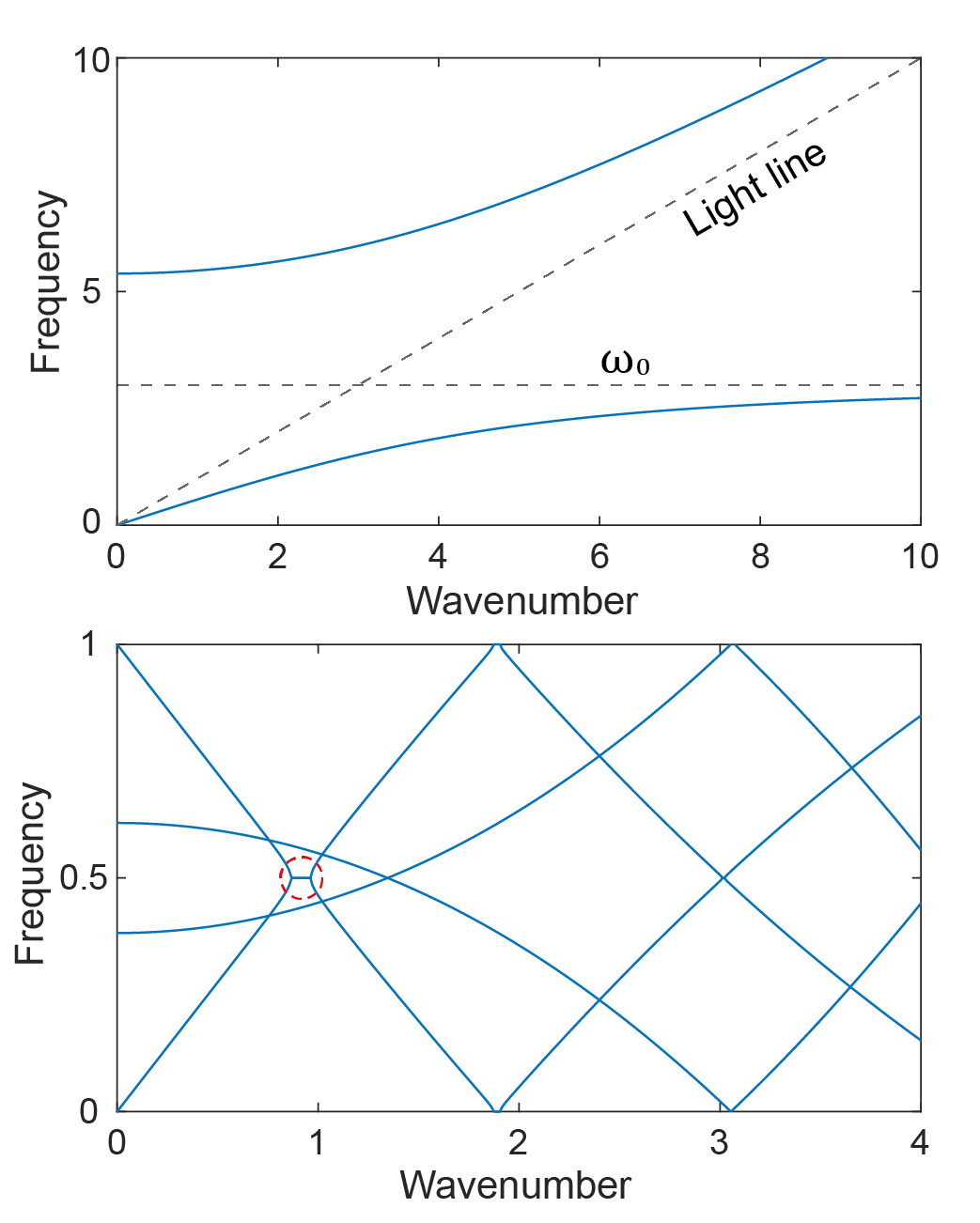}
  \caption{\label{fig:sup_LZ_band}
The band structure derived from an undriven Lorentz model (upper panel) is shown alongside the Floquet band structure from a driven Lorentz model (lower panel). In the driven Lorentz model, the elastic constant is modeled as $\kappa(t) = \kappa_0 + \kappa_m \sin(\Omega t)$. The band structures are plotted using the following fictitious parameters: $\kappa_0 = 9$, $\kappa_m = 1.8$, $N = 20$, $m = 1$, $e = -1$, and $\Omega = 1$. Additionally, the vacuum permittivity and permeability are assumed to be 1.}
\end{figure}
\begin{figure}[htb!]
  \centering
    \includegraphics[width=0.45\textwidth]{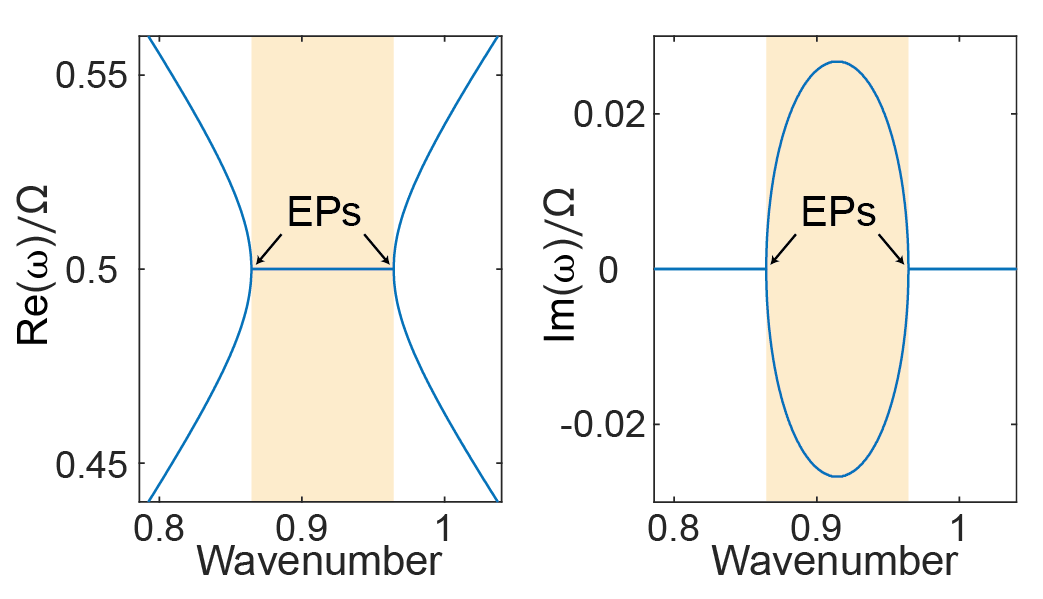}
  \caption{\label{fig:sup_MG}Illustration of the Floquet band structure near the momentum gap of the PTC as described by the driven Lorentz model. The plots are generated under the assumption of a lossless case, where $\eta = 0$.}
\end{figure}
When the elastic constant is modulated in a time-periodic manner, the field vector solutions to the Schrödinger-like equation take the form of a Floquet mode, i.e., $\ket{\Psi(t)} = e^{-i\omega t}\ket{\Phi(t)}$. Here, $\ket{\Phi(t)}$ exhibits the same periodicity as the Hamiltonian matrix $\mathcal{H}(t)$, enabling the Fourier expansion of $\ket{\Phi(t)}$ as $\ket{\Phi(t)} = \sum_n{e^{-in\Omega t}\ket{\phi_n}}$. Additionally, the time-periodic Hamiltonian matrix can be expanded as $\mathcal{H}(t) = \sum_m{e^{-im\Omega t}\mathcal{H}_m}$. Substituting the expanded Hamiltonian matrix and field vectors into the Schrödinger-like equation reveals that:
\begin{equation}
    (\omega+n\Omega)\ket{\phi_n}= 
\begin{cases}
    \sum_m{\mathcal{H}_{n-m}\ket{\phi_m}}+\ket{s},& \text{if } n=0\\
    \sum_m{\mathcal{H}_{n-m}\ket{\phi_m}}, & \text{otherwise}
\end{cases} \label{eqn:ifeq} 
\end{equation}
where $n$ and $m$ are integers. The above system of linear equations can be recast into the following matrix equation similar to that given in the previous section:
\begin{equation}
    \omega\ket{F}=\mathcal{H}_F\ket{F}+\ket{s_F}, \label{eqn:LZ_Floquet_matrix_Eq} 
\end{equation}
where $\ket{s_F}$ is a column vector defined as $\ket{s_F} = [\cdots; 0; \ket{s}; 0; \cdots]$, and the Floquet Hamiltonian matrix, $\mathcal{H}_F$, is expressed as
\begin{equation}
    \mathcal{H}_F = \begin{bmatrix}
        \ddots & \ddots & & & \\
        \ddots & \mathcal{H}_0+\Omega I & \mathcal{H}_{-1} & \mathcal{H}_{-2} & \\
               & \mathcal{H}_{+1} & \mathcal{H}_0 & \mathcal{H}_{-1} & \\
               &    \mathcal{H}_{+2}         & \mathcal{H}_{+1} & \mathcal{H}_0-\Omega I & \ddots \\
               &    &   & \ddots & \ddots
    \end{bmatrix}. \label{eqn:Floquet_Ham} 
\end{equation}

The Floquet band structure, characterized by complex-valued quasi-eigenfrequencies, can be calculated by solving the following eigenvalue problem under a source-free condition ($J=0$):
\begin{equation}
    \mathcal{H}_F\ket{R}=\omega\ket{R}. \label{eqn:HF_eig} 
\end{equation}
When employing a driven Lorentz model with a sinusoidally varying elastic constant, the matrices $\mathcal{H}_{\pm m}$, which describe the coupling between the original bands and the Floquet sidebands, are zero for all non-zero integer values of $m$, except for $m=1$. This indicates that only adjacent bands interact. As illustrated in the bottom panel of Figure \ref{fig:sup_LZ_band}, the Floquet band structure reveals the emergence of a momentum gap, highlighted by the red dashed circle. This momentum gap emerges at the intersection of the original positive frequency band and the first-order Floquet sideband of the negative frequency band. Figure \ref{fig:sup_MG} illustrates the Floquet band structure in the vicinity of this momentum gap.

\begin{figure}[htb!]
  \centering
    \includegraphics[width=0.45\textwidth]{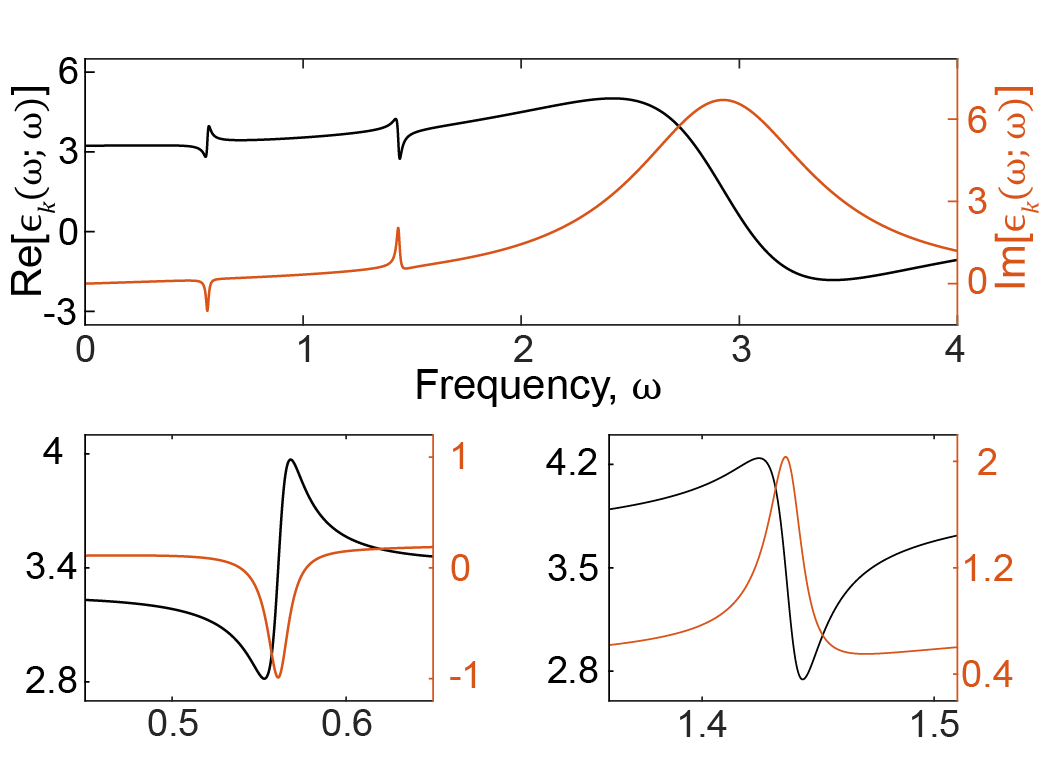}
  \caption{\label{fig:sup_LZ_eps}The complex-valued relative permittivity in the case of small intrinsic loss: $\eta = 1 < \eta_c$. For this plot, the wavenumber is fixed at $k = 0.8$.}
\end{figure}

Next, we investigate the low-loss case, characterized by $\eta = 1$, which is less than the critical value $\eta_c$. By applying Eq. \eqref{eqn:LZ_Floquet_matrix_Eq}, we calculate the complex-valued, momentum-resolved relative permittivity, defined as $\epsilon(k,\omega) = 1 + P_{0}/E_0$, and present it in Fig. \ref{fig:sup_LZ_eps} for a fixed wavenumber $k = 0.8$. Besides the fundamental Lorentzian resonance at $\omega_0 = \sqrt{\kappa_0/m}$, two additional Lorentzian-like resonances emerge due to the time-periodic modulation, as specifically illustrated in Fig. \ref{fig:sup_LZ_eps}. The resonance at the lower frequency is attributed to the Floquet sideband of the negative frequency band, whereas the resonance at the higher frequency is related to the Floquet sideband of the positive frequency band. Notably, near the lower frequency resonance, the imaginary part of the relative permittivity becomes negative, indicating net gain. This region of net gain corresponds to the negative $\mathrm{kDOS}$ region identified through power flow analysis. Figure \ref{fig:sup_LZ_kDOS} illustrates the maps of $\bar{P}_{D}(k,\omega)$ and $\bar{P}_{E}(k,\omega)$ for a case of small intrinsic loss. The SE decay and excitation rates, shown in the left panels, demonstrate significant qualitative agreement with the results derived from the nondispersive model discussed in the main text.
\clearpage
\begin{figure}[htb!]
  \centering
    \includegraphics[width=0.9\textwidth]{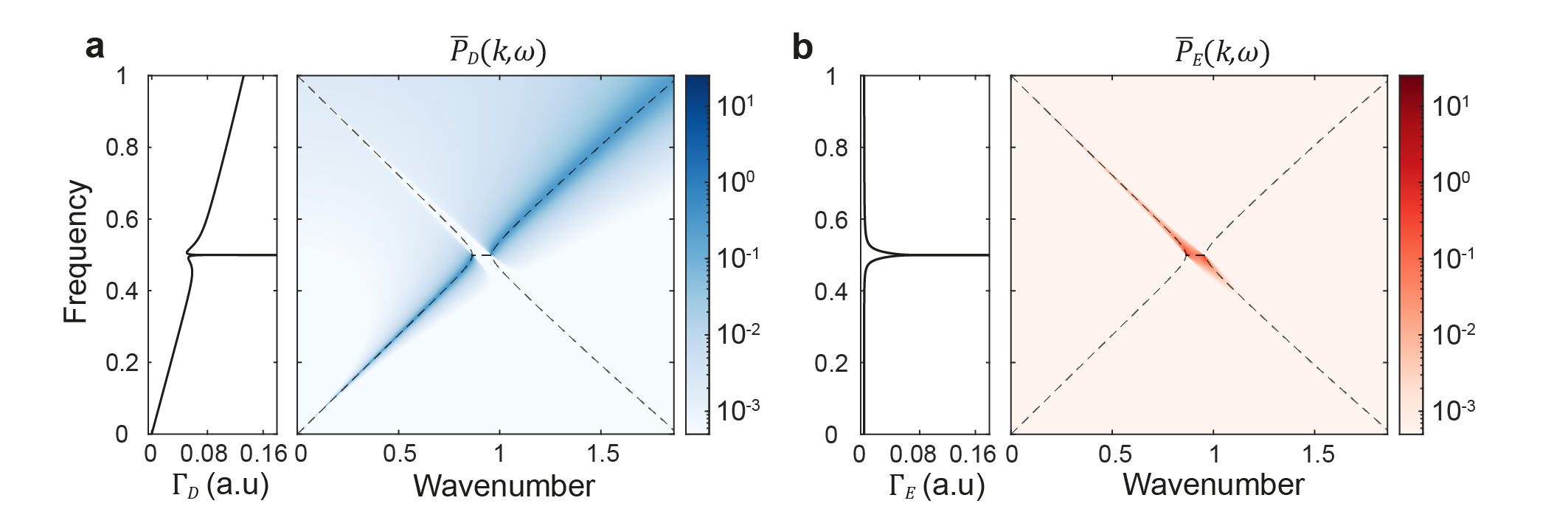}
  \caption{\label{fig:sup_LZ_kDOS}Maps of $\bar{P}_{D}(k,\omega)$ and $\bar{P}_{E}(k,\omega)$, along with SE decay and excitation rates, for $\eta = 1$ (indicating a case of small intrinsic loss).}
\end{figure}

\section{$\mathrm{kDOS}$ decomposition within momentum gap}
In this section, we focus on the decomposition of $\mathrm{kDOS}$ when the wavenumber is located within the central gap region ( $k_{EP_-}<\mathbf{k}<k_{EP_+}$). When considering only the two bands involved in the formation of the momentum gap, it is no longer valid to assume that $\mathrm{Im}[I_\alpha(\mathbf{k})]$ is negligible for this range of wavenumbers, where
\begin{equation}
    I_\alpha (\mathbf{k}) = \mathbf{n}_\mathbf{p}\cdot\left[2\delta_0\cdot\frac{\ket{R_\alpha(\mathbf{k})}\bra{L_\alpha(\mathbf{k})}}{\braket{L_\alpha(\mathbf{k})}{R_\alpha(\mathbf{k})}}\cdot \epsilon_0A_F\otimes\mathbb{I}\right]\cdot\mathbf{n}_\mathbf{p}.
\end{equation} 
Consequently, the $\mathrm{kDOS}$ is expressed as follows:
\begin{equation}
    \rho_\mathbf{p}(\mathbf{k},\omega) \approx\sum_{\alpha=\pm}\rho_{\mathbf{p},\mathrm{Re}}^\alpha(\mathbf{k},\omega)+\rho_{\mathbf{p},\mathrm{Im}}^\alpha(\mathbf{k},\omega),
\end{equation}
where two constituting $\rho_\mathbf{p}(\mathbf{k},\omega)$ terms are given by,
\begin{equation}
\begin{aligned}
       &\rho_{\mathbf{p},\mathrm{Re}}^\alpha(\mathbf{k},\omega)=\frac{1}{\pi}\frac{\gamma_\alpha}{(\omega-\Omega_\alpha)^2+\gamma_\alpha^2}\mathrm{Re}[I_\alpha (\mathbf{k})],\\
       &\rho_{\mathbf{p},\mathrm{Im}}^\alpha(\mathbf{k},\omega)=\frac{1}{\pi}\frac{\Omega_\alpha-\omega}{(\omega-\Omega_\alpha)^2+\gamma_\alpha^2}\mathrm{Im}[I_\alpha (\mathbf{k})].
\end{aligned}
\end{equation}
Figure \ref{fig:sup_kDOS_MG} shows the $\mathrm{kDOS}$ and its decomposition for a case with low intrinsic loss, specifically when $\sigma=0.1$. 
\begin{figure}[h!]
  \centering
    \includegraphics[width=0.45\textwidth]{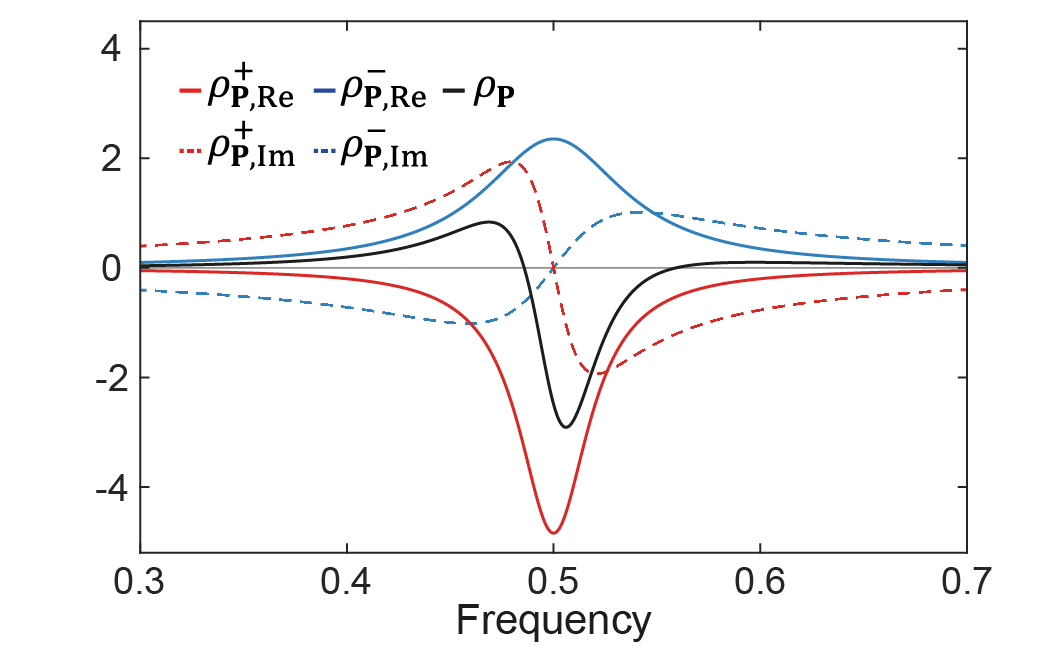}
  \caption{\label{fig:sup_kDOS_MG}$\mathrm{kDOS}$ and its decomposition plotted as a function of frequency at $k=1.05$ for the case of low intrinsic loss, i.e., $\sigma=0.1<\sigma_c$.}
\end{figure}
\section{Non-orthogonality of Floquet eigenmodes}
Due to the non-Hermiticity of the Floquet Hamiltonian matrix, the Floquet eigenmodes $\ket{\Phi_m^\alpha(t)}$ are generally no longer orthogonal. Here, the integer $m$ denotes Floquet index and $\alpha=\pm$ label the bands according to their propagation directions.  Inside the momentum gap, the bands are indistinguishable by propagation direction. Therefore, we label the band with a smaller imaginary part of the eigenfrequency as $\alpha=+$. The {\it extended} inner product between two Floquet eigenmodes are defined as,
\begin{equation}
    \langle\langle\Phi_m^\alpha(t)\vert\Phi_n^\beta(t)\rangle\rangle
    = \frac{1}{T}\int_0^T{dt \braket{\Phi_m^\alpha(t)}{\Phi_n^\beta(t)}}
\end{equation}
The integrand can be expressed as,
\begin{align}
    &\braket{\Phi_m^\alpha(t)}{\Phi_n^\beta(t)}\nonumber\\
    &=\left(\sum_p{e^{i(p-m)\Omega t}\bra{\phi_p^\alpha}}\right)\left(\sum_p{e^{-i(p-n)\Omega t}\ket{\phi_p^\beta}}\right)\nonumber\\
    &=\sum_p{\sum_q{e^{-i(p-2q-m+n)\Omega t}\braket{\phi_{p-q}^\alpha}{\phi_q^\beta}}}
\end{align}
After time-averaging, only the terms with $p-2q-m+n=0$ remain, yielding
\begin{align}
    \langle\langle\Phi_m^\alpha(t)\vert\Phi_n^\beta(t)\rangle\rangle &= \sum_q{\braket{\phi_{q+m-n}^\alpha}{\phi_q^\beta}}\\
    &=\sum_q{\braket{\phi_{q+m}^\alpha}{\phi_{q+n}^\beta}}\label{eqn:innerprod1}
\end{align}
The inner product of the Floquet (right) eigenvectors, on the other hand, can be expressed as,
\begin{align}
    \braket{R_m^\alpha}{R_n^\beta}=\sum_q{\braket{\phi_{q+m}^\alpha}{\phi_{q+n}^\beta}}
\end{align}
As a result, the inner product of the Floquet right eigenvectors is equivalent to the suitably extended inner product of the Floquet eigenmodes. Because the Floquet Hamiltonian matrix is non-Hermitian, the Floquet eigenmodes of PTCs are generally non-orthogonal to each other when the wavenumber is fixed. The Petermann factor ($\mathrm{PF}$) is a measure of non-orthogonality in a non-Hermitian system that was originally introduced to quantify the excess noise induced by the non-orthogonal resonant modes of unstable cavities\cite{1070064,Berry2003,Zhang2018}. In our case, the $\mathrm{PF}$ can be written as, 
\begin{equation}
    \mathrm{PF} = \frac{\braket{R_m^\alpha}{R_m^\alpha}\braket{L_m^\alpha}{L_m^\alpha}}{\left|\braket{L_m^\alpha}{R_m^\alpha}\right|^2} \label{eqn:def_PF}
\end{equation}
where the Floquet left eigenvectors $\bra{L_m^\alpha}$ are defined as the solutions of the following eigenvalue equation,
\begin{equation}
    \bra{L_m^\alpha}\mathcal{H}_F=\bra{L_m^\alpha}\omega_m^\alpha
\end{equation}
Figure \ref{fig:sup_PF} shows the $\mathrm{PF}$ for the band indicated by the red arrow in the inset calculated from Eq.\eqref{eqn:def_PF}. The $\mathrm{PF}$ diverges as the wavenumber approaches the gap edges because of the vanishing overlap between Floquet left and right eigenvectors $\braket{L_m^\alpha}{R_m^\alpha}$.
\clearpage
\begin{figure}[htb!]
  \centering
    \includegraphics[width=0.45\textwidth]{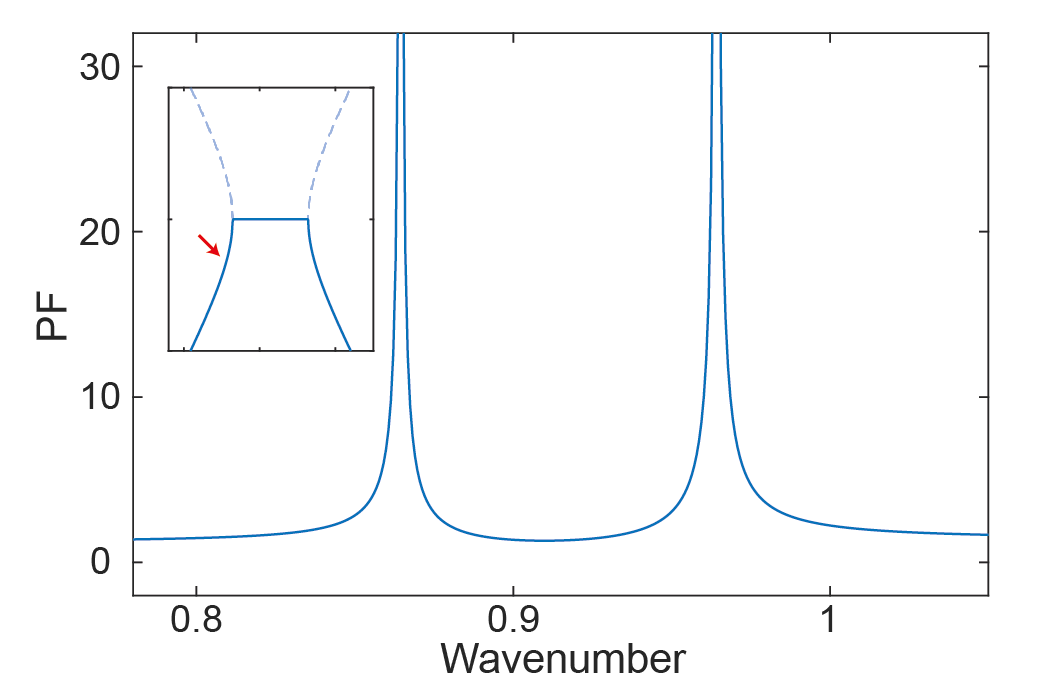}
  \caption{\label{fig:sup_PF}PF calculated along the band indicated by the red arrow in the inset}
\end{figure}

\section{$\mathrm{kDOS}$ at EPs}
In this section, we follow the methodology outlined in \cite{Pick2017} to derive the $\mathrm{kDOS}$ at the edge of the momentum gap. Precisely at $\mathbf{k_{EP_+}}$ (i.e., the right edge of the momentum gap), the Floquet Hamiltonian matrix assumes the form of a defective matrix, which we denote as $\mathcal{H}_F^0$. The right Floquet eigenvector and its corresponding right Jordan vector at the exceptional point (EP) can be described by

\begin{equation}
    \begin{aligned}
    \mathcal{H}_F^{0}\ket{R^0}&=\omega_0\ket{R^0}\\
    \mathcal{H}_F^0\ket{J_R^0}&=\omega_0\ket{J_R^0}+\ket{R^0}.     
    \end{aligned}
     \label{eqn:ERandJR}
\end{equation}

Here, $\omega_0=\Omega_0-i\gamma_0$ denotes the degenerate quasi-eigenfrequency at the EP. In the vicinity of $\mathbf{k_{EP_+}}$ (specifically, $|\mathbf{k}|=k_{EP_+} + \lambda$, where $\lambda\ll 1$), the Floquet Hamiltonian matrix is expanded as $\mathcal{H}_F(\lambda)=\mathcal{H}_F^0+\lambda\mathcal{H}_F^1$. The eigenvalue equation for this near-EP regime is given by:
\begin{equation}
    \mathcal{H}_F(\lambda)\ket{R^\pm}=\omega_\pm \ket{R^\pm}.
    \label{eqn:eigP_near_EP}
\end{equation}
Near the second-order EP, it is posited that both the quasi-eigenfrequencies and the right Floquet eigenvectors can be expressed using alternating Puiseux series, as described in \cite{Pick2017}:
\begin{equation}
\begin{aligned}
    \omega_\pm&=\omega_0\pm \lambda^{1/2}\omega_1+\lambda\omega_2 \pm \lambda^{3/2}\omega_3+\cdots \\
    \ket{R^\pm}&=\ket{R^0}\pm \lambda^{1/2}\ket{R^1}+\lambda\ket{R^2}\pm \cdots.
\end{aligned}
\label{eqn:eig_near_EP}
\end{equation}
By inserting Eq. \eqref{eqn:eig_near_EP} into Eq. \eqref{eqn:eigP_near_EP} and matching the coefficients of $\lambda^{1/2}$ and $\lambda$, we infer that
\begin{equation}
\begin{aligned}
    \omega_\pm&=\omega_0\pm \lambda^{1/2}\Delta+\mathcal{O}(\lambda)\\
    \ket{R^\pm}&=\ket{R^0}\pm \lambda^{1/2}\Delta\ket{J_R^0}+\mathcal{O}(\lambda),
\end{aligned}
\label{eqn:eig_near_EP2}
\end{equation}
where $\Delta = (\bra{L^0}\mathcal{H}_F^1\ket{R^0}/\braket{J_L^0}{R^0})^{1/2}$. The same analysis can be applied to the left Floquet vector and the left Jordan vector. Utilizing Eq. \eqref{eqn:eig_near_EP2}, the Green's function at $\mathbf{k_{EP_+}}$ can be calculated as follows:
\begin{equation}\label{eq:green's_function_at_EP_lossless}
\begin{aligned}
        \mathcal{G}_F(\mathbf{k_{EP_+}},\omega)&\approx\lim_{\lambda\rightarrow0}\left[\sum_{\alpha=\pm}\frac{1}{\omega-\omega_\alpha}\frac{\ket{R^\alpha}\bra{L^\alpha}}{\braket{L^\alpha}{R^\alpha}}\right]=\\
        &\frac{2}{(\omega-\omega_0)^2}\frac{\ket{R^0}\bra{L^0}}{\braket{L^0}{J_R^0}+\braket{J_L^0}{R^0}}
        +\frac{2}{\omega-\omega_0}\frac{\ket{R^0}\bra{J_L^0}+\ket{J_R^0}\bra{L^0}}{\braket{L^0}{J_R^0}+\braket{J_L^0}{R^0}}.
\end{aligned}
\end{equation}
Then, $\rho_\mathbf{p}(\mathbf{k_{EP_+}},\omega)$ is expressed as follows:
\begin{equation}\label{eq:green's_function_at_EP_lossy}
\begin{aligned}
    \rho_\mathbf{p}(\mathbf{k_{EP_+}},\omega)&\approx\frac{2\epsilon_0\mu\omega}{\pi }\mathrm{Im}[\mathbf{n}_\mathbf{p}\cdot\mathbb{G}_0(\mathbf{k_{EP_+}},\omega)\cdot\mathbf{n}_\mathbf{p}]\\
    &\approx\frac{2}{\pi}\frac{\{\gamma_0^2-(\omega-\Omega_0)^2\}}{\{(\omega-\Omega_0)^2+\gamma_0^2\}^2}\mathrm{Im}\left[\mathbf{n}_\mathbf{p}\cdot\left\{2\delta_0\cdot\frac{\ket{R^0}\bra{L^0}}{\braket{L^0}{J_R^0}+\braket{J_L^0}{R^0}}\cdot\epsilon_0A_F\otimes\mathbb{I}\right\}\cdot\mathbf{n}_\mathbf{p}\right]\\
    &+\frac{4}{\pi}\frac{\gamma_0(\omega-\Omega_0)}{\{(\omega-\Omega_0)^2+\gamma_0^2\}^2}\mathrm{Re}\left[\mathbf{n}_\mathbf{p}\cdot\left\{2\delta_0\cdot\frac{\ket{R^0}\bra{L^0}}{\braket{L^0}{J_R^0}+\braket{J_L^0}{R^0}}\cdot\epsilon_0A_F\otimes\mathbb{I}\right\}\cdot\mathbf{n}_\mathbf{p}\right]\\
    &+\frac{2}{\pi}\frac{\Omega_0-\omega}{(\omega-\Omega_0)^2+\gamma_0^2}\mathrm{Im}\left[\mathbf{n}_\mathbf{p}\cdot\left\{2\delta_0\cdot\frac{(\ket{R^0}\bra{J_L^0}+\ket{J_R^0}\bra{L^0})}{\braket{L^0}{J_R^0}+\braket{J_L^0}{R^0}}\cdot\epsilon_0A_F\otimes\mathbb{I}\right\}\cdot\mathbf{n}_\mathbf{p}\right]\\
    &+\frac{2}{\pi}\frac{\gamma_0}{(\omega-\Omega_0)^2+\gamma_0^2}\mathrm{Re}\left[\mathbf{n}_\mathbf{p}\cdot\left\{2\delta_0\cdot\frac{(\ket{R^0}\bra{J_L^0}+\ket{J_R^0}\bra{L^0})}{\braket{L^0}{J_R^0}+\braket{J_L^0}{R^0}}\cdot\epsilon_0A_F\otimes\mathbb{I}\right\}\cdot\mathbf{n}_\mathbf{p}\right].
\end{aligned}
\end{equation}

\clearpage
\end{widetext}

\end{document}